\documentclass[10pt, a4paper, onecolumn, twoside]{article}

\usepackage{graphicx, xcolor}

\usepackage[top=35.2truemm,bottom=35.2truemm,left=34.5truemm,right=34.5truemm]{geometry}

\title{\bf \Large 
Hierarchically Modular Dynamical Neural Network \\ 
Relaxing in a Warped Space: \\
Basic Model and its Characteristics 
}

\author{
Kazuyoshi Tsutsumi
\thanks{
Department of Mechanical Engineering and Robotics, 
Department of Mechanical and Systems Engineering, 
Ryukoku University, 
Otsu-shi, Shiga-ken 520-2194, Japan}
\and 
Ernst Niebur
\thanks{The Zanvyl Krieger Mind/Brain Institute, 
Johns Hopkins University, 
3400 N. Charles Street, Baltimore, MD 21218, USA} 
}

\date{}

\begin{document}

\maketitle 
\thispagestyle{empty}

\begin{abstract} 
\noindent
We propose a hierarchically modular, dynamical neural network model 
whose architecture minimizes a specifically designed energy function 
and defines its temporal characteristics. 
The model has an internal and an external space 
that are connected with a layered ``internetwork" 
that consists of a pair of forward and backward subnets 
composed of static neurons (with an instantaneous time-course). 
Dynamical neurons with large time constants in the internal space 
determine the overall time-course. 
The model offers a framework in which state variables 
in the network relax in a warped space, 
due to the cooperation between dynamic and static neurons. 
We assume that the system operates 
in either a learning or an association mode, 
depending on the presence or absence of feedback paths and input ports. 
In the learning mode, synaptic weights in the internetwork 
are modified by strong inputs 
corresponding to repetitive neuronal bursting, 
which represents sinusoidal or quasi-sinusoidal waves 
in the short-term average density of nerve impulses  
or in the membrane potential. 
A two-dimensional mapping relationship can be formed 
by employing signals with different frequencies 
based on the same mechanism as Lissajous curves. 
In the association mode, the speed of convergence to a goal point 
greatly varies with the mapping relationship 
of the previously trained internetwork, 
and owing to this property, 
the convergence trajectory in the two-dimensional model 
with the non-linear mapping internetwork 
cannot go straight but instead must curve. 
We further introduce a constrained association mode 
with a given target trajectory 
and elucidate that in the internal space, an output trajectory is generated, 
which is mapped from the external space 
according to the ``inverse" of the mapping relationship 
of the forward subnet. 

\vspace{2mm}
\noindent
{\bf Keywords:} hierarchical; modular; dynamical neural network; neuronal bursting; 
inverse mapping; Lissajous 
% 
%%%%% ----- Updated Part No. 00 (Begin) ----- 
%%%%% v1 
% curves. 
%%%%% v2 
curve. 
%%%%% ----- Updated Part No. 00 (End) ----- 
% 
\end{abstract}

\markboth{\rm Relaxing in a Warped Space}{\rm Kazuyoshi Tsutsumi \& Ernst Niebur}

% 1 Introduction
\section{Introduction}

\noindent
One of the earliest models of neurons was proposed 
by McCulloch and Pitts
\cite{McCulloch1943}.
In combination with Hebb's insights 
into neuronal mechanisms of learning
\cite{Hebb1949},
their static neuron model was the basis of such well-known learning machines 
as the Perceptron, ADALINE, and the Learning Matrix
\cite{Rosenblatt1958}\cite{Widrow1960}\cite{Steinbuch1963}. 
The period from the late 1950s to the early 1960s is significant 
in the sense that neural network research started in earnest
\cite{Block1962}.
An important limitation that surfaced about a decade later is 
that the XOR problem cannot be learned in the original Perceptron architecture
\cite{Minsky1969}.
Consequently, a quiet period in neural network research followed, 
although multiple interesting ideas and models 
were proposed even during this era
\cite{Amari1967}\cite{Amari1977}\cite{Nakano1972}
\cite{Kohonen1972}\cite{Werbos1974}\cite{Grossberg1976}.
In 1986, the generalized delta rule was presented 
with specific simulation examples 
as an advanced version of the original delta rule, 
allowing multi-layered networks with non-linear neurons to be trained
\cite{Rumelhart1986a}\cite{Rumelhart1986b},
and this revitalized neural network research. 
During this period, expectations about a ``neurocomputer'' also arose
\cite{Hect-Nielsen1988}.
Although neural network-based computation may have failed to 
reach the level of practical use that had been expected by some, 
the 1980s can be considered a crucial period 
because of the large number of novel frameworks proposed in this time
\cite{Kohonen1982}\cite{Hecht-Nielsen1987}.

In parallel with studies of artificial neural networks, 
attempts to understand the function of biological neural networks 
based on the identified circuitry 
of the brain have a long history. 
Following the pioneering studies of the visual system by Hubel and Wiesel
\cite{Hubel1959},
the Cognitron and Neocognitron were introduced as computational models 
with a highly hierachical structure
\cite{Fukushima1975}\cite{Fukushima1980}\cite{Fukushima1983}.
Even though it currently remains unclear whether 
the gap between biological knowledge and artificial models is narrowing, 
interest in deeply hierarchical neural structures continues to grow, 
and 
research into how to train such layered networks 
has been receiving attention in recent years. 
The excellent performances 
obtained in image and voice recognition
\cite{Hinton2006a}\cite{Hinton2006b}\cite{Markoff2012}~
can be pointed out as the reason for this growing interest. 
Hierarchy and modularity in neural architecture appear 
to be significant organizing principles 
that may effectively and adequately reduce the informational dimension
\cite{Ozawa2009}\cite{Hafner2017}.
Based on mathematical analyses of human neuroimaging data, research 
into searching modular and/or hierarchical structures from the viewpoint 
of understanding their organizing principles has attracted increasing attention
\cite{Meunier2010}\cite{Nicolini2016}.

Research related to the dynamics of neurons and networks 
originated with the modeling of neural membranes 
and with the analysis of circuits 
based primarily on a comparatively small number of neurons. 
The first deep understanding of the biophysics of neuronal membranes 
came with a mathematical model of the squid giant axon proposed in 1952
\cite{Hodgkin1952}.
The complexity of this model was reduced to the phenomena 
considered essential in a set of lower-dimensional models
\cite{FitzHugh1955}\cite{Nagumo1962}.
With regards to the circuit analysis, 
the dynamical characteristics of a neuron 
are generally accounted for, but in most cases, 
are simplified to their essentials. 
Circuits are typically constructed 
with inhibitory and excitatory connections, and occasionally gap junctions
\cite{Niebur1988};
their dynamical behavior is examined 
in terms of the mean frequency fluctuation of nerve impulses, 
which can be further abstracted 
with respect to oscillation (limit cycles), chaos, etc.
\cite{Morishita1972}\cite{Aihara1999}.
~Such approaches allow us to simplify the network dynamics 
sufficiently to understand, for instance, the behavior of large networks 
whose dynamics are proposed to stay in or close to a critical state
\cite{Millman2010}\cite{Plenz2014}.
These issues are further discussed 
in terms of synaptic plasticity and neuronal adaptation 
that offer the fascinating power and capability of neural networks
\cite{Tsutsumi1984a}\cite{Tsutsumi1984b}
\cite{Tsutsumi1988a}\cite{Tsutsumi1988b}\cite{Matsuoka1985}.
Specifically, the importance and effectiveness of 
Central Pattern Generators (CPGs) based on relatively small networks 
with dynamical neurons is being recognized 
for controlling locomotive robots
\cite{Tsujita2003}.
In studies on dynamical neurons, 
either a biologically identified network 
or a network supposedly constructed in a building block manner 
is often employed. 
After a new idea was proposed, 
where the superordinate concept called an energy function 
can be utilized for deriving a network architecture
\cite{Hopfield1982}\cite{Hopfield1984}, 
a study of generic dynamical neural networks 
was conducted using large-scaled networks 
\cite{Carpenter1987}\cite{Tsutsumi1987}\cite{Tsutsumi1988b}\cite{Tsutsumi1988c}. 
This framework's proposal was indispensable 
as a factor to increase the expectations 
for the above-mentioned neurocomputer, 
particularly with regard to parallel computation. 

One overarching theme of signaling routes 
in the brain and the nervous system 
is the bidirectionality in connectivity. 
Even if we focus on a local area, 
mutual communication exists between layers, modules, and/or sub-modules. 
One example of models with mutual connections between layers 
is described in a series of those called Adaptive Resonance Theory
\cite{Grossberg2013}.
On a larger scale, reciprocal connections between layers 
are also essential for generating organism-wide functions, 
like selective attention
\cite{Fukushima1986}. 
Thus, understanding network architectures in which 
multiple layers are connected reciprocally and in a multiplexing manner 
is a matter of great interest. 

Of particular interest are studies in which various approaches 
from different fields are associated with one another. 
For instance, learning mechanisms and activity dynamics 
have frequently been discussed separately: 
the first in the context of how synaptic plasticity and static mapping 
could be related, and the second in understanding how activity 
varies in a dynamical neural network with fixed synaptic connections. 
In this sense, a series of models with bidirectional connections
named Bidirectional Associative Memories (BAM) was highly influential
\cite{Kosko1988},
and training algorithms called Back-Propagation Through Time 
and Real-Time Recurrent Learning, 
by which a dynamical mapping relationship can be acquired 
in a recurrent neural network, have great potential
\cite{Williams1990}\cite{Williams1995}. 
In addtion, it is important to model effects that can be produced 
by combining multiple types of neurons with different time constants
\cite{Tsutsumi1990}\cite{Tsutsumi1991}\cite{Ozawa1998}. 
To clarify the existence of higher mechanisms in the brain, 
multiple fundamental principles must be integrated 
for a more general framework. 
One challenge is 
understanding the behavior of a dynamical neural network 
in which the characteristics of various neurotransmitters 
are accurately reflected
\cite{Doya2002}. 
Even though constructing a model is difficult 
using only common knowledge, 
it might be reasonable to build a consistent hypothesis
\cite{Fukushima2013}. 

In this paper, we try to construct a model 
that combines multiple frameworks as 
cooperative dynamical responses 
by two types of neurons with different time constants, 
fluctuations of neuronal firing rates including bursting activity, 
as well as hierarchy and modularity in neural architecture, 
bidirectional (centripetal and centrifugal) paths, 
and mapping with generalization. 
We then propose a hierarchically modular neural network model 
that operates using cooperativity between dynamic and static neurons. 
This model is based on two previous studies. The first 
\cite{Tsutsumi1989} 
uses an architecture in which dynamical neurons are sandwiched 
between two static layered neural networks 
with mutually inverse mapping relationships, 
and it includes local and global feedback loops. 
Its distinctive feature is that, 
in a computational task to be processed, 
each of the different constraints can be satisfied 
conveniently in a warped inner space or a linear outer space 
owing to a mapper with an inverse mapping relationship that is ``explicitly" set. 
However, that model was derived according to a bottom-up design approach, 
and it lacked the theoretical grounding that we propose in the current study 
in terms of the temporal derivative of an energy function. 
The other study
\cite{Tsutsumi2003} 
is a predecessor of our model, 
but it was limited to a simple proposal that assumed only a linear mapper. 
Furthermore, its dynamical behavior including how to train 
the total network was not discussed. 

In Section 2, we describe the model and its derivation process in detail. 
Following a clarification of the network architecture, 
we explain the two kinds of operating modes: learning and association. 
In Section 3, we investigate the learning mode, i.e., how a part of the whole network, 
in which synaptic plasticity is supposed, can be trained 
in collaboration with activity dynamics. 
Starting from simulation studies of 
a one-dimensional model with linear mapping, 
we refer to a model with non-linear mapping 
and then expand it into a two-dimensional model. 
In the following section, we examine, as an association mode, 
the dynamical behavior of the network trained in the previous section. 
In Section 5, on the basis of the simulation results in Sections 3 and 4, 
we discuss the relationship between the learning and association modes 
as well as experiments that precisely investigate 
network dynamics as a constrained association mode, 
in which a target trajectory (instead of a goal point) 
is applied to an input port of the model. 
This work is curiosity-driven fundamental research 
\cite{Flexner2017}
and was not developed with a view on direct applicability for solving practical problems. 
We summarize our results and look toward future work in the last section.

% 2 Model
\section{Model}

% 2.1 Basic Concept
\subsection{Basic Concept}

\noindent
The following is an example of a dynamical neural network equation:

% Equation-01
\vspace{1mm} 
\begin{equation} 
- c_{i} \frac{dU_{i}}{dt} = 
\sum_{j} T_{ij} V_{j} - \sum_{k} S_{ik} Z_{k} 
- J_{i} + \frac{U_{i}}{r_{i}} ~.
\end{equation}

\vspace{1mm} 
\noindent 
\( U_{i} \) and \( V_{i} \) are 
the inner state (membrane potential) of the \( i \)-th neuron 
and its output voltage (short-time average impulse density) respectively; 
both are time-varying functions 
that are related to each other 
with the following input-output function:

% Equation-02
\vspace{1mm} 
\begin{equation}
V_{i} = g(~U_{i}~) ~.
\end{equation}

\vspace{1mm} 
\noindent
\( T_{ij} \) means the fixed-valued conductance of a direct connection 
from the \( j \)-th dynamical neuron to the \( i \)-th one.
\( Z_{k} \) is the voltage (short-term average impulse density) 
of an input from the outside of the model 
that is added to a time-constant circuit composed of dynamical neurons 
after being converted from voltage to current by the conductance \( S_{ik} \).
\( J_{i} \) is a bias current injected from the outside. 
\( c_{i} \) and \( r_{i} \) are the values of capacitance and resistance 
in the \( i \)-th dynamical neuron, 
which together determine its time constant. 
The changes of the sign of each term in Eq. (1) 
determine whether the dynamics is convergent or divergent; 
this issue has often been discussed 
in terms of the energy function stated below 
after the framework was first proposed by Hopfield. 

Consider the following energy function:

% Equation-03
\vspace{0mm} 
\begin{eqnarray} 
E & \stackrel{\triangle}{=} & 
\frac{1}{2} \sum_{i} V_{i} ~( ~\sum_{j} T_{ij} V_{j} ~)
~- \sum_{i} V_{i} ~( ~\sum_{k} S_{ik} Z_{k} ~)
\nonumber \\ 
  &                         & 
\hspace{5mm}
- \sum_{i} V_{i} J_{i} 
~+ \sum_{i} \frac{1}{r_{i}} \int_{0}^{V_{i}} g^{-1}(V) dV ~.
\end{eqnarray}

\vspace{2mm} 
\noindent 
Differentiating this energy function with respect to time, we get

% Equation-04
\vspace*{1mm} 
\begin{equation} 
\frac{dE}{dt} = \sum_{i} \frac{dV_{i}}{dt} 
\mbox{\LARGE [} 
\sum_{j} T_{ij} V_{j} 
- \sum_{k} S_{ik} Z_{k} 
- J_{i} + \frac{U_{i}}{r_{i}} 
\mbox{\LARGE ]} ~.
\end{equation}

\vspace{1mm} 
\noindent 
Since the terms in the brackets of the right-hand side of Eq. (4) 
are the same as the right-hand side of Eq. (1), 
we can substitute Eq. (1) for Eq. (4). 
If we employ, in the network described by Eq. (1), 
the following monotonically increasing function 
as \( g(x) \) in Eq. (2)

% Equation-05
\vspace{0mm} 
\begin{equation} 
g(x) ~=~ \alpha x
~~~~~~~~~~( ~\alpha > 0 ~: ~\mbox{Const.} ~) ~, 
\end{equation}

\vspace{2mm} 
\noindent 
or

% Equation-06
\vspace{0mm} 
\begin{equation} 
g(x) ~=~ \frac{1}{1 + e^{- \alpha x}}
~~~~~~~~~~( ~\alpha > 0 ~: ~\mbox{Const.} ~) ~, 
\end{equation}

\vspace{2mm} 
\noindent 
taking \( g'(~U_{i}~) > 0 \) into consideration, 
application of the chain rule in Eq. (4) yields

% Equation-07
\vspace{0mm} 
\begin{eqnarray} 
\frac{dE}{dt} & = & - ~\sum_{i} ~c_{i} 
                    ~\mbox{\Large (} \frac{dV_{i}}{dt} \mbox{\Large )} 
                    ~\mbox{\Large (} \frac{dU_{i}}{dt} \mbox{\Large )} 
                    \nonumber \\ 
              & = & - ~\sum_{i} ~c_{i} ~g^{\prime} (U_{i}) 
                    ~\mbox{\Large (} \frac{dU_{i}}{dt} \mbox{\Large )}^{2} 
                    ~~\leq 0 ~. 
\end{eqnarray}

\vspace{2mm} 
\noindent 
We thus find that \( dE / dt \) is non-positive, 
showing that the value of the energy function 
given by Eq. (3) always decreases toward a global or a local minimum. 
By defining a task to be processed, such as an associative memory 
or an optimization problem, in the form of an energy function like Eq. (3), 
various problem-solving methods based on the dynamical neural network 
given by Eqs. (1) and (6) have been developed previously 
\cite{Hopfield1985}\cite{Tank1986}. 
It should be noted that dynamics described by Eq. (1) 
generally include both convergent 
( \( U_{i} \rightarrow a \) / \( V_{i} \rightarrow b \) ) 
and divergent 
( ( \( U_{i} \rightarrow - \infty \) / \( V_{i} \rightarrow 0 \) ) or 
( \( U_{i} \rightarrow + \infty \) / \( V_{i} \rightarrow 1 \) ) ) behaviors, 
where \( a \) and \( b \) are finite constants. 

We will use a superscript with the form of \( < \hspace{-1mm} 0 \hspace{-1mm} > \) 
to the output of a dynamical neuron \( V_{i} \) 
to specify that neurons exist in the \( 0 \)-th space (the internal space). 
In addition, we suppose that 
the variables \( V^{<0>}_{i} \), \( (i = 0, 1, 2, ...) \) 
are converted to the output ones 
\( V^{<1>}_{i^{\prime}}, ~( ~i^{\prime} = 0, 1, 2, ... ~) \) 
in the upper \( 1 \)-st space (the external space) 
based on the following function:

% Equation-08
\vspace{1mm}
\begin{equation}
V^{<1>}_{i^{\prime}} = f^{<1>}_{i^{\prime}}
( V^{<0>}_{0}, V^{<0>}_{1}, V^{<0>}_{2}, ... ~) ~.
\end{equation}

\vspace{2mm} 
\noindent 
Using the converted variables \( V^{<1>}_{i^{\prime}} \), 
we now re-define the total energy function given by Eq. (9)
on the assumption that the direct fixed-valued connection 
\( T^{<P>}_{ij} \), ( \( P=0,1 \) ) is symmetric:

% Equation-09
\vspace{-2mm}
\begin{eqnarray}
E & \stackrel{\triangle}{=} &
    \frac{1}{2} \sum_{i^{\prime}} V^{<1>}_{i^{\prime}}
    ~( ~\sum_{j} T^{<1>}_{i^{\prime}j} ~V^{<1>}_{j} ~)
    ~- ~\sum_{i^{\prime}} V^{<1>}_{i^{\prime}}
    ~( ~\sum_{k} S^{<1>}_{i^{\prime}k} ~Z^{<1>}_{k} ~)
    \nonumber \\
  &                         &
    + ~\frac{1}{2} \sum_{i} V^{<0>}_{i}
    ~( ~\sum_{j} T^{<0>}_{ij} ~V^{<0>}_{j} ~)
    ~- ~\sum_{i} V^{<0>}_{i}
    ~( ~\sum_{k} S^{<0>}_{ik} ~Z^{<0>}_{k} ~)
    \nonumber \\
  &                         &
    + ~\sum_{i} \frac{1}{r_{i}}
    \int_{0}^{V^{<0>}_{i}} g^{-1}(V) dV ~.
\end{eqnarray}

\vspace{1mm} 
\noindent 
Differentiating Eq. (9) with respect to time, we obtain

% Equation-10
\vspace{-2mm} 
\begin{eqnarray} 
\frac{dE}{dt} & = & \sum_{i^{\prime}} \frac{dV^{<1>}_{i^{\prime}}}{dt} 
                    \mbox{\Large (} 
                    \sum_{j} T^{<1>}_{i^{\prime}j} V^{<1>}_{j}
                    - \sum_{k} S^{<1>}_{i^{\prime}k} Z^{<1>}_{k}
                    \mbox{\Large )} 
                    \nonumber \\ 
              &   & + \sum_{i} \frac{dV^{<0>}_{i}}{dt} 
                    \mbox{\Large (}
                    \sum_{j} T^{<0>}_{ij} V^{<0>}_{j} 
                    - \sum_{k} S^{<0>}_{ik} ~Z^{<0>}_{k} 
                    + \frac{U_{i}}{r_{i}} 
                    \mbox{\Large )} ~. 
\end{eqnarray}

\vspace{1mm} 
\noindent 
\( dV^{<1>}_{i^{\prime}} / dt \) in Eq. (10) 
can be transformed as follows:

% Equation-11
\vspace{0mm}
\begin{equation}
\frac{d V^{<1>}_{i^{\prime}}}{dt} ~=~ \sum_{i}
\frac{\partial V^{<1>}_{i^{\prime}}}{\partial V^{<0>}_{i}}
\frac{d V^{<0>}_{i}}{dt} ~.
\end{equation}

\vspace{1mm} 
\noindent 
Substituting Eq. (11) for Eq. (10), we get

% Equation-12
\vspace{-2mm}
\begin{eqnarray}
\frac{dE}{dt} & = & \sum_{i} \frac{dV^{<0>}_{i}}{dt}
                    ~\mbox{\LARGE [}
                    ~\sum_{i^{\prime}}
                    \frac{\partial V^{<1>}_{i^{\prime}}}{\partial V^{<0>}_{i}}
                    \mbox{\Large (}
                    ~\sum_{j} T^{<1>}_{i^{\prime}j} V^{<1>}_{j}
                    ~- ~\sum_{k} S^{<1>}_{i^{\prime}k} Z^{<1>}_{k}
                    ~\mbox{\Large )}~
                    \nonumber \\
              &   & + \sum_{j} T^{<0>}_{ij} V^{<0>}_{j}
                    - \sum_{k} S^{<0>}_{ik} ~Z^{<0>}_{k}
                    + \frac{U_{i}}{r_{i}}
                    ~\mbox{\LARGE ]} ~.
\end{eqnarray}

\vspace{1mm} 
\noindent 
If we construct a dynamical neural network described by the following equation,

% Equation-13
\vspace{0mm} 
\begin{eqnarray} 
- ~c_{i} \frac{dU_{i}}{dt} 
              & = & \sum_{j} T^{<0>}_{ij} V^{<0>}_{j} 
                    - \sum_{k} S^{<0>}_{ik} ~Z^{<0>}_{k} 
                    \nonumber \\ 
              &   & \hspace{-3mm} 
                    + \sum_{i^{\prime}} 
                    ~\frac{\partial V^{<1>}_{i^{\prime}}}{\partial V^{<0>}_{i}}
                    ~\mbox{\Large (}
                    ~\sum_{j} T^{<1>}_{i^{\prime}j} V^{<1>}_{j} 
                    ~- \sum_{k} S^{<1>}_{i^{\prime}k} Z^{<1>}_{k} 
                    ~\mbox{\Large )}
                    + \frac{U_{i}}{r_{i}} ~, 
\end{eqnarray}

\vspace{2mm} 
\noindent 
it always holds that \( dE/dt \leq 0 \) 
in the same way as in Eq. (7), and therefore the value of the energy function 
defined by Eq. (9) monotonically decreases as time proceeds. 
Figure 1 shows the block diagram of the ladder-shaped network 
described by Eq. (13). 
The part surrounded by a gray dotted line 
corresponds to the network represented by Eq. (1). 
Here the conversion from \( V^{<0>}_{i} \) to \( V^{<1>}_{i^{\prime}} \) 
by Eq. (8) is called {\sl Mapper M}, and the product-sum operation with 
 \( \partial V^{<1>}_{i^{\prime}} / \partial V^{<0>}_{i} \) 
is called {\sl Mapper N}. 
The network has two kinds of feedback paths. 
One is in the internal space and is included in Eq. (1). 
The other is in the external space, 
and due to this feedback path 
by {\sl Mapper M}, the connections \( T^{<1>}_{i^{\prime}j} \), and {\sl Mapper N}, 
the output signal in the internal space 
is returned again to the dynamical neurons in the internal space.

% Figure 01
\begin{figure}[!t]

    \hspace*{-6.5mm}
    \includegraphics[scale=0.90]{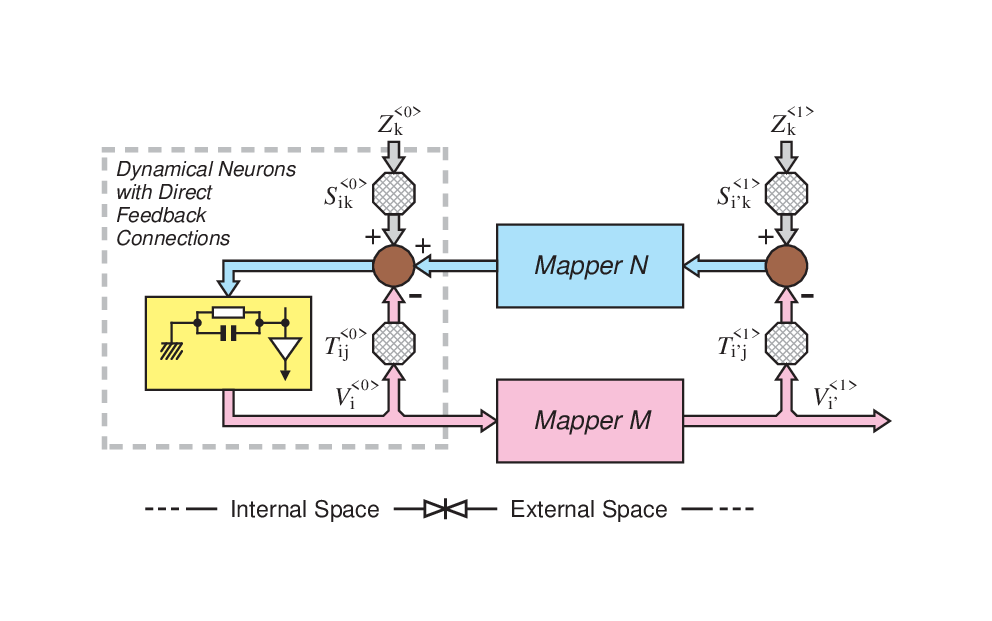}

    \vspace*{-8.0mm}
    \caption{
      A block diagram of the network described by Eq. (13). 
      Conversion from \( V^{<0>}_{i} \) to \( V^{<1>}_{i^{\prime}} \) 
      is called {\sl Mapper M}, and a product-sum operation with 
      \( \partial V^{<1>}_{i'} / \partial V^{<0>}_{i} \) is called {\sl Mapper N}. 
      The network has two kinds of feedback paths. 
      One is in the internal space and is included in Eq. (1). 
      The other is via the external space; 
      the output in the internal space is returned back to dynamical neurons 
      in the internal space through {\sl Mapper M}, 
      the connections \( T^{<1>}_{i^{\prime}j} \), 
      and {\sl Mapper N}. 
      The network also has an input port in each space. 
    } 

\end{figure}

% 2.2 Static Neurons
\subsection{Static Neurons}

\noindent
In the previous subsection, 
we assumed that \( V^{<0>}_{i} \) and \( V^{<1>}_{i^{\prime}} \) 
were related to each other by Eq. (8). 
Now we use a layered neural network composed of neurons 
with a very small time constant that can be regarded as static. 
In the formulation, the layer number is assigned from \( 0 \) to \( L-1 \), 
and the sum of the inputs applied to each neuron in the \( l \)-th layer 
and the neuronal output are 
\( net^{<1><l>}_{i} \) and \( O^{<1><l>}_{i} \). 
A modifiable connection from the \( j \)-th neuron in the \( l \)-th layer 
to the \( i \)-th neuron in the \( (l + 1) \)-th layer 
is written as \( W^{<1><l>}_{ij} \). 
We regard the output of a neuron in the layered network \( O^{<1><L-1>}_{i} \) 
as the output signal in the external space \( V^{<1>}_{i} \) 
and maintain consistency with Eq. (8). 
The first superscript with the brackets \( < \) and \( > \) 
in such variables as \( O \), \( net \), and \( W \) 
is \( < \hspace{-0.75mm} 1 \hspace{-0.75mm} > \), 
since it is the one that becomes necessary 
after establishing the external space. 
Note that we redundantly employ 
such indices as \( i \), \( j \), and \( k \) 
without their maximum values 
to reduce the number of indices. 

A neuron in the input layer of the layered network 
is assumed to have linear characteristics represented by Eq. (5), 
and we make the input layer a simple relaying function 
to send \( V^{<0>}_{i} \) directly to the layered network.
In general, we employ a non-linear neuron described by Eq. (6) 
in the hidden layers, and a linear or non-linear neuron in the output layer. 
For simplicity, we assume that each neuron shares linearity or non-linearity 
inside the corresponding layer, 
although each can have its own input-output characteristics. 
The relation between these variables is as follows:

% Equation-14 & Equation-15
\vspace{-2mm} 
\begin{eqnarray} 
V^{<1>}_{i}    & \stackrel{\triangle}{=} & O^{<1><L-1>}_{i} \\
               &                         & \nonumber \\
O^{<1><l>}_{i} & =                       & g ( ~net^{<1><l>}_{i} ~) 
\end{eqnarray}

\vspace{0mm} 
\[
\mbox{where~~~~~}
net^{<1><l>}_{i} = \sum_{j} W^{<1><l-1>}_{ij} O^{<1><l-1>}_{j} ~. 
\]

\vspace{-3mm}
\[
\hspace{25mm}
( ~l = L-1, L-2, ..., 2, 1 ~)
\]

\vspace{1mm} 
\noindent 
The following is the time derivative of Eq. (14):

% Equation-16
\vspace{-2mm}
\begin{eqnarray}
\frac{d V^{<1>}_{i}}{dt} & = &
        g^{\prime} ( ~net^{<1><L-1>}_{i} ~)
        ~\times ~\mbox{\Large (}
        ~\sum_{j} \frac{d W^{<1><L-2>}_{ij}}{dt} ~O^{<1><L-2>}_{j}
        \nonumber \\
                         &   &
        \hspace{20mm}
        + \sum_{j} W^{<1><L-2>}_{ij} ~\frac{d O^{<1><L-2>}_{j}}{dt}
        ~\mbox{\Large )} ~.
\end{eqnarray}

\vspace{1mm} 
\noindent 
Let us define \( H \) as the first term 
on the right-hand side of Eq. (10),

% Equation-17
\vspace{-2mm}
\begin{eqnarray}
H & \stackrel{\triangle}{=} &
    \sum_{i} \frac{d V^{<1>}_{i}}{dt}
    ~\mbox{\Large (}
    ~\sum_{j} T^{<1>}_{ij} V^{<1>}_{j}
    ~- \sum_{k} S^{<1>}_{ik} Z^{<1>}_{k}
    ~\mbox{\Large )} ~,
\end{eqnarray}

\vspace{1mm} 
\noindent 
and substitute Eq. (16) for Eq. (17). 
Then, after some rearrangements, we obtain

% Equation-18
\vspace{-2mm} 
\begin{eqnarray} 
H & = & \sum_{i} \sum_{j} \frac{d W^{<1><L-2>}_{ij}}{dt}
        ~O^{<1><L-2>}_{j} ~\delta^{<1><L-1>}_{i}
        \nonumber \\ 
  &   & \hspace{0mm} 
        ~+ \sum_{j} \frac{d O^{<1><L-2>}_{j}}{dt}
        ~\times ~\mbox{\Large (} 
        ~\sum_{i} W^{<1><L-2>}_{ij} ~\delta^{<1><L-1>}_{i}
        ~\mbox{\Large )} ~.
\end{eqnarray}

\vspace{1mm}
\noindent
Introducing \( \delta^{<1><L>}_{i} \) for convenience,

% Equation-19
\vspace{-2mm}
\begin{eqnarray}
\delta^{<1><L>}_{i}   & \stackrel{\triangle}{=} &
        \sum_{j} T^{<1>}_{ij} V^{<1>}_{j}
        - \sum_{k} S^{<1>}_{ik} Z^{<1>}_{k} ~, 
\end{eqnarray}

\vspace{1mm}
\noindent
\( \delta^{<1><L-1>}_{i} \) in Eq. (18) is written as follows:

% Equation-20
\vspace{-2mm}
\begin{eqnarray}
\delta^{<1><L-1>}_{i} & \stackrel{\triangle}{=} &
        g^{\prime} ( ~net^{<1><L-1>}_{i} ~) ~\delta^{<1><L>}_{i} ~.
\end{eqnarray}

\vspace{1mm} 
\noindent 
The variable \( \delta \) 
is also needed after the supposition of the external space, 
so the first superscript with \( < \) and \( > \) is assumed 
to be \( < \hspace{-0.75mm} 1 \hspace{-0.75mm} > \).
Further calculating \( d O^{<1><L-2>}_{j} / dt \) in Eq. (18) 
based on Eq. (15) yields

% Equation-21
\vspace{1mm} 
\begin{eqnarray} 
H & = & \sum_{i} \sum_{j} \frac{d W^{<1><L-2>}_{ij}}{dt}
        ~O^{<1><L-2>}_{j} ~\delta^{<1><L-1>}_{i}
        \nonumber \\ 
  &   & \hspace{-2mm} 
        + ~\sum_{j} \sum_{k} \frac{d W^{<1><L-3>}_{jk}}{dt}
        ~O^{<1><L-3>}_{k} ~\delta^{<1><L-2>}_{j}
        \nonumber \\ 
  &   & \hspace{-2mm} 
        + ~\sum_{k} \frac{d O^{<1><L-3>}_{k}}{dt}
        ~\times
        ~\mbox{\Large (} 
        ~\sum_{j} W^{<1><L-3>}_{jk} ~\delta^{<1><L-2>}_{j}
        ~\mbox{\Large )} ~,
\end{eqnarray}

\vspace{1mm} 
\noindent 
where

% Equation-22
\vspace{-2mm} 
\begin{eqnarray} 
~\delta^{<1><L-2>}_{j} & \stackrel{\triangle}{=} & 
g^{\prime} ( ~net^{<1><L-2>}_{j} ~)
~\times 
~\mbox{\Large (}
~\sum_{i} W^{<1><L-2>}_{ij} \delta^{<1><L-1>}_{i} 
~\mbox{\Large )} ~. 
\end{eqnarray}

\vspace{1mm} 
\noindent 
We obtain the following equation after repeating these calculations 
from the output layer to the input layer:

% Equation-23
\vspace{-2mm}
\begin{eqnarray}
H & = & \sum_{l=0}^{L-2}
        ~\mbox{\LARGE [}
        ~\sum_{i} \sum_{j} \frac{d W^{<1><l>}_{ij}}{dt}
        ~\times 
        ~\mbox{\Large (}
        ~O^{<1><l>}_{j} ~\delta^{<1><l+1>}_{i}
        ~\mbox{\Large )}
        ~\mbox{\LARGE ]}
        \nonumber \\
  &   & ~+ \sum_{i} \frac{d O^{<1><0>}_{i}}{dt}
        ~\mbox{\Large (}
        \sum_{j} W^{<1><0>}_{ji} ~\delta^{<1><1>}_{j}
        \mbox{\Large )} ~. 
\end{eqnarray}

\vspace{1mm} 
\noindent 
Here,  \( \delta \) is written with Eqs. (20) and (22) as follows:

% Equation-24
\vspace{-2mm} 
\begin{eqnarray} 
\delta^{<1><l>}_{i} & = & g^{\prime} ( ~net^{<1><l>}_{i} ~)
                          ~\times 
                          ~\mbox{\Large (}
                          ~\sum_{j} W^{<1><l>}_{ji} \delta^{<1><l+1>}_{j} 
                          ~\mbox{\Large )} ~. 
                          \nonumber \\
                    &   & \hspace{20mm}
                          ( ~l = L-1, L-2, ..., 2, 1 ~)
\end{eqnarray}

\vspace{1mm} 
\noindent 
% Old Version
% Linear neurons are assumed in the input layer of the Internetwork 
% and simply relay the output of the dynamical neurons in the internal space 
% to the upper layer of {\sl Forward Subnet}; 
% so we can put 
% 
% New Version 2021.02.23 Ten-nou Tanjou-bi
As for the layered network, linear neurons are assumed in the input layer 
and simply relay the output of the dynamical neurons in the internal space 
to the upper layer; so we can put

% Equation-25
\vspace{1mm} 
\begin{equation} 
O^{<1><0>}_{i} = V^{<0>}_{i} ~.
\end{equation}

\vspace{2mm} 
\noindent 
Therefore, returning Eqs. (23) and (24) to Eq. (10) based on the above, 
we finally obtain

% Equation-26
\vspace{-2mm} 
\begin{eqnarray} 
\frac{dE}{dt} & = & \sum_{i} \frac{d V^{<0>}_{i}}{dt}
                    ~\mbox{\Large (} 
                    ~\sum_{j} T^{<0>}_{ij} V^{<0>}_{j} 
                    ~- \sum_{k} S^{<0>}_{ik} Z^{<0>}_{k}
                    ~+ \delta^{<1><0>}_{i} 
                    ~+ \frac{U_{i}}{r_{i}} 
                    ~\mbox{\Large )}
                    \nonumber \\ 
              &   & \hspace{0mm} 
                    + \sum_{l=0}^{L-2} 
                    ~\mbox{\LARGE [} 
                    ~\sum_{i} \sum_{j} \frac{d W^{<1><l>}_{ij}}{dt}
                    ~\times 
                    ~\mbox{\Large (} 
                    ~O^{<1><l>}_{j} ~\delta^{<1><l+1>}_{i} 
                    ~\mbox{\Large )} 
                    ~\mbox{\LARGE ]} ~,
\end{eqnarray}

\vspace{2mm} 
\noindent 
where

% Equation-27
\begin{equation} 
\delta^{<1><0>}_{i} ~\stackrel{\triangle}{=} 
~\sum_{j} W^{<1><0>}_{ji} \delta^{<1><1>}_{j} ~.
\end{equation}

\vspace{1mm} 
\noindent 
Here, constructing the network described by Eq. (28) 
and assuming that the connections in the layered neural network are modified 
according to Eq. (29),

% Equation-28 
\vspace{-2mm} 
\begin{eqnarray} 
- ~c_{i} ~\frac{dU_{i}}{dt}               & = & 
        \sum_{j} T^{<0>}_{ij} V^{<0>}_{j} 
        - \sum_{k} S^{<0>}_{ik} Z^{<0>}_{k} 
        + \delta^{<1><0>}_{i} 
        + \frac{U_{i}}{r_{i}} ~, 
\end{eqnarray}

% Equation-29
\vspace{-5mm} 
\begin{eqnarray} 
- \eta^{<1><l>}_{ij} \frac{d W^{<1><l>}_{ij}}{dt} & = & 
        O^{<1><l>}_{j} ~\delta^{<1><l+1>}_{i} ~, 
        \\
                                          &   & 
        \hspace{10mm} 
        ( ~l = L-2, L-3, ..., 1, 0 ~)
        \nonumber 
\end{eqnarray}

\vspace{1mm} 
\noindent 
\( dE/dt \leq 0 \) always holds in the same way as in Eq. (7), 
and therefore the value of the energy function defined by Eq. (9) 
also monotonically decreases as time proceeds. 
\( \eta^{<1><l>}_{ij} \) is a positive constant determining the learning speed. 
\( \delta \) is given by Eqs. (19), (24), and (27), 
and \( \delta^{<1><0>}_{i} \) in Eq. (27) can be calculated 
from \( \delta^{<1><L>}_{i} \) according to Eqs. (19) and (24). 

Expressing the computational process of \( \delta \) explicitely 
as network architecture, 
the total network described by Eqs. (28) and (29) is illustrated 
as a block diagram in Fig. 2. 
Obviously, by comparison with Fig. 1, 
{\sl Mapper M} and {\sl Mapper N} in Fig. 1 
are replaced by two layered networks 
that are complementary to each other. 
In the following sections, we call such a pair of layered networks 
{\sl Internetwork} in the sense that it connects two spaces, 
and we name these two layered networks 
{\sl Forward Subnet} and {\sl Backward Subnet} respectively. 
In Fig. 2, we also call {\sl Internetwork}'s left part the internal space 
and its right part the external space.

% Figure 02
\begin{figure}[!t]

    \hspace*{-6.5mm}
    \includegraphics[scale=0.90]{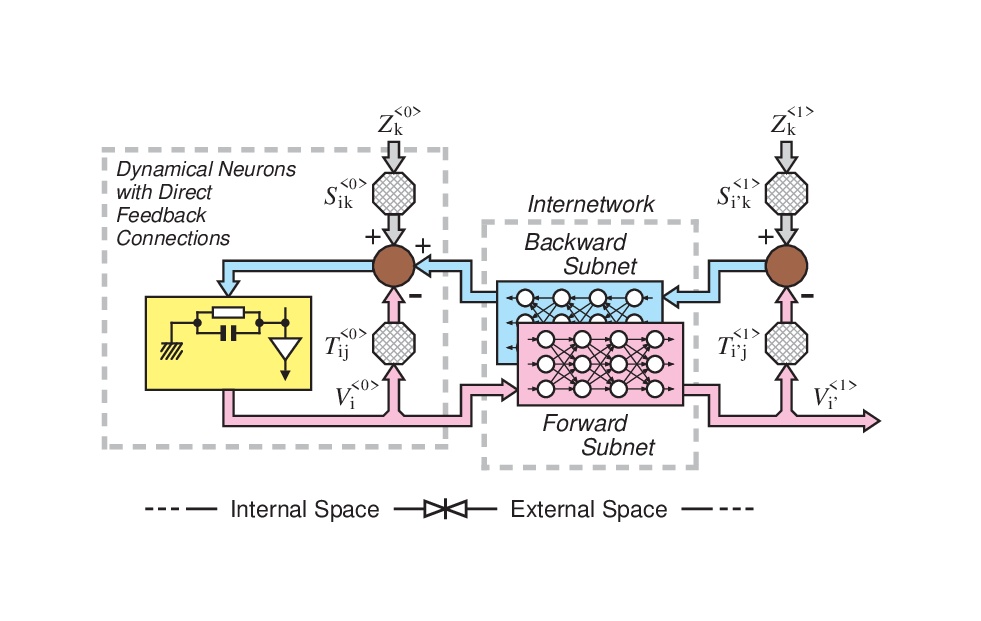}

    \vspace*{-8.0mm}
    \caption{
      Architecture of the proposed network described by 
      Eqs. (28) and (29), Eqs. (14) and (15), and Eqs. (19), (24), and (27). 
      The network has a pair of a feedback path and an input port in each space. 
      {\sl Mapper M} and {\sl Mapper N} shown in Fig. 1 
      are replaced by a pair of layered networks 
      that are complementary to each other. 
      The computational process of \( \delta \) is expressed 
      explicitly as a layered network. 
    } 

\end{figure}

% 2.3 Learning and Association Modes 
\subsection{Learning and Association Modes}

\noindent
As shown in Fig. 2, our model has one feedback path and one input port 
in both the internal and external spaces. 
Changing the values of \( T_{ij} \) and \( S_{ik} \) and their signs 
in the first and second terms 
on the right-hand side of the energy function Eq. (3), 
the shape of the potential field can be chosen arbitrarily. 
It is reasonable, in this sense, that 
\( T_{ij} \) and \( S_{ik} \) are treated as a pair. 
In the architecture shown by Figs. 1 or 2, 
\( T^{<0>}_{ij} \) and \( S^{<0>}_{ik} \) 
exist in the internal space and 
\( T^{<1>}_{i^{\prime}j} \) and \( S^{<1>}_{i^{\prime}k} \) 
in the external space.
For instance, consider a situation where 
\( T^{<0>}_{ij} \) and \( S^{<0>}_{ik} \) 
in the internal space are removed. 
Then, since there is no inner feedback loop based on 
\( T^{<0>}_{ij} \) and \( S^{<0>}_{ik} \), 
the network dynamics are determined only by 
a detoured feedback loop with {\sl Forward Subnet} (or {\sl Mapper M}), 
\( T^{<1>}_{i^{\prime}j} \) and \( S^{<1>}_{i^{\prime}k} \), and 
{\sl Backward Subnet} (or {\sl Mapper N}).
How does the network behave in such a limited feedback loop? 
It is quite interesting to investigate its time-course characteristics. 
Therefore, prior to studying the model's behavior in the following sections, 
we group the model into three topological types as illustrated in Fig. 3, 
depending on the presence or absence of connections 
in the internal and external spaces. 
The upper one is the case with all the connections, 
and we call it the Learning Mode. 
The middle and lower types are called the Association Mode; 
the middle one (Type \#0) is the case in which 
a pair of \( T \) and \( S \) only exists in the internal space, 
and the lower one (Type \#1) is the case in which 
it only exists in the external space.

% Figure 03
\begin{figure}[!t]

    \vspace*{-3.0mm}
    \begin{center}

      \includegraphics[scale=0.90]{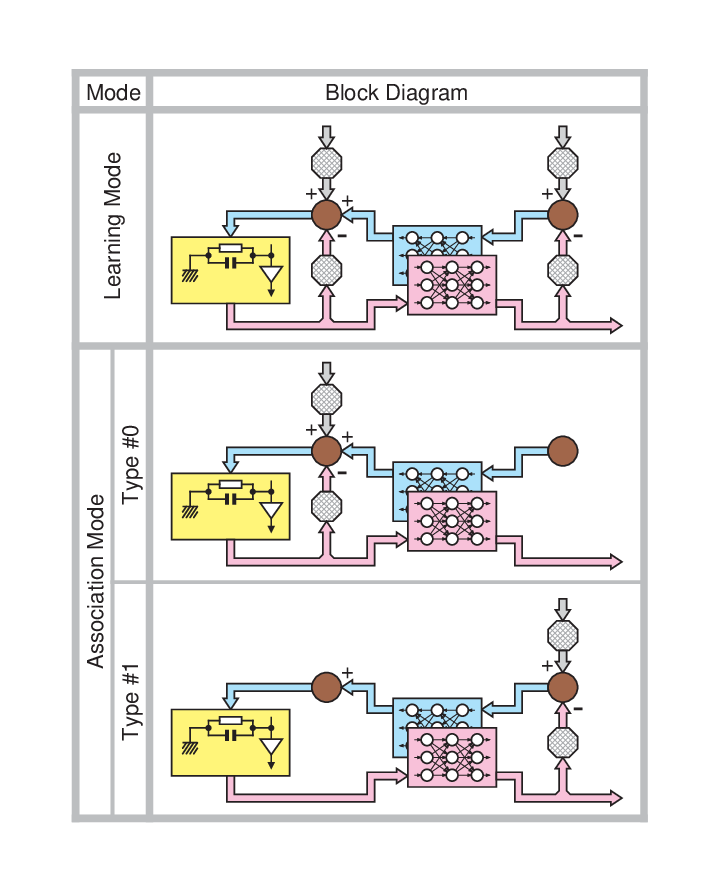}

    \end{center}

    \vspace*{-10.0mm}
    \caption{
      Basically, the model is topologically grouped 
      into two types, totally three types, 
      depending on the presence or absence of connections 
      in the internal and external spaces.
      The upper row shows the case with all of feedback paths and input ports, 
      and we call it ``Learning Mode." 
      The middle and lower types are called ``Association Mode"; 
      the middle one shows the case in which 
      a pair of \( T \) and \( S \) exists 
      only in the internal space (Type \#0), 
      and the lower one does the case in which 
      it only exists in the external space (Type \#1). 
    } 

\end{figure}

In the Learning Mode architecture, 
both an input signal and a target (teaching) signal necessary for learning 
can be applied to {\sl Internetwork}. 
In the Association Mode architectures, on the contrary, 
either an input signal or a target (teaching) signal 
appears to be given to {\sl Internetwork}, 
so the synaptic connections should not properly be modified. 
Therefore, it is reasonable in avoiding this situation 
that we assume a condition in relation to the learining dynamics 
given by Eq. (29) in our proposed network positioned as a model, 
when we examine how the activity dynamics differ 
according to such architectures as Types \#0 and \#1 in Fig. 3. 
Thus, the following condition in Eq. (29) is presumed 
in conjunction with such a topological distinction; 
Eq. (29) is always ON in the Learning Mode, 
but it is OFF in the Association Mode, 
assuming the existence of a certain mechanism. 
The activity dynamics described by Eq. (28) 
is ON at all times in both modes.

% 3 Learning Mode
\section{Learning Mode}

% 3.1 Learning Mode in One-Dimensional Model 
\subsection{Learning Mode in One-Dimensional Model}

\vspace{1mm} 
\noindent 
(1) Linear Mapping

\vspace{2mm} 
\noindent 
In the network architecture derived in the previous section, 
only synaptic connections in {\sl Internetwork} were assumed to have plasticity. 
Eq. (28) is for the activity dynamics of the whole network, 
and Eq. (29) is for the learning dynamics 
on the plastic connections in {\sl Internetwork}; 
these two kinds of dynamics run at all times 
in the network of the Learning Mode. 
Therefore, we do not need any other particular learning algorithm 
even when the training issues are discussed for the model, 
and we should only trace the time-course behavior of the total network 
when appropriate signals are applied to the input ports. 

What kind of signals can we put to those input ports from the outside? 
For {\sl Internetwork} to acquire a wide-ranging mapping relationship, 
we can easily imagine that signals with fixed values are inadequate 
as inputs in the internal and external spaces. 
Although we can consider various types of signals, 
we use a sinusoidal wave as an example of the input signal. 
Since \( V^{<p>}_{i} \) and \( Z^{<p>}_{i}, ~(p = 0, 1) \) 
are assumed to be of short-time average impulse density, 
the sinusoidal wave here practically means 
bursting or semi-bursting nerve impulses. 
In our first experiment, 
we applied identical sinusoidal waves 
to the two input ports in the internal and external spaces 
of the whole network in Fig. 2, in which 
the activity dynamics work according to Eq. (28) 
and the learning dynamics in {\sl Internetwork} with modifiable synapses 
run based on Eq. (29). 
We assume that a dynamical neuron in the internal space 
has linear input-output characteristics 
given by Eq. (5) with \( \alpha = 1.0 \). 
We conduct our simulation studies 
based on the standard Runge-Kutta method 
with a stepsize of \( 0.1 ~{\sf ms} \) 
which was determined in preliminary studies 
to yield the correct solution of the equations. 
We set the values of \( c_{i} \) and \( \eta_{ij} \) 
so that the time constants of the activity and learning dynamics 
are \( 1 ~{\sf ms} \) and \( 5 ~{\sf s} \). 
\( r_{i} \) is set to \( \infty \). 
Then, the frequency of both input signals 
applied in the internal and external spaces 
is set to \( 10.0 ~{\sf Hz} \). 
Each of {\sl Forward Subnet} and {\sl Backward Subnet} 
is assumed to be a three-layered network; 
we use, in the input and output layers, linear neurons 
whose characteristics are given by Eq. (5) with \( \alpha = 1.0 \), 
and in the hidden layer, non-linear ones 
represented by Eq. (6) with \( \alpha = 10.0 \). 
In a one-dimensional model, 
we employ eight hidden neurons in {\sl Internetwork}.

% Figure 04(a)(b)
\begin{figure}[!t]

   \hspace*{-6.0mm} 
   \includegraphics[scale=0.72]{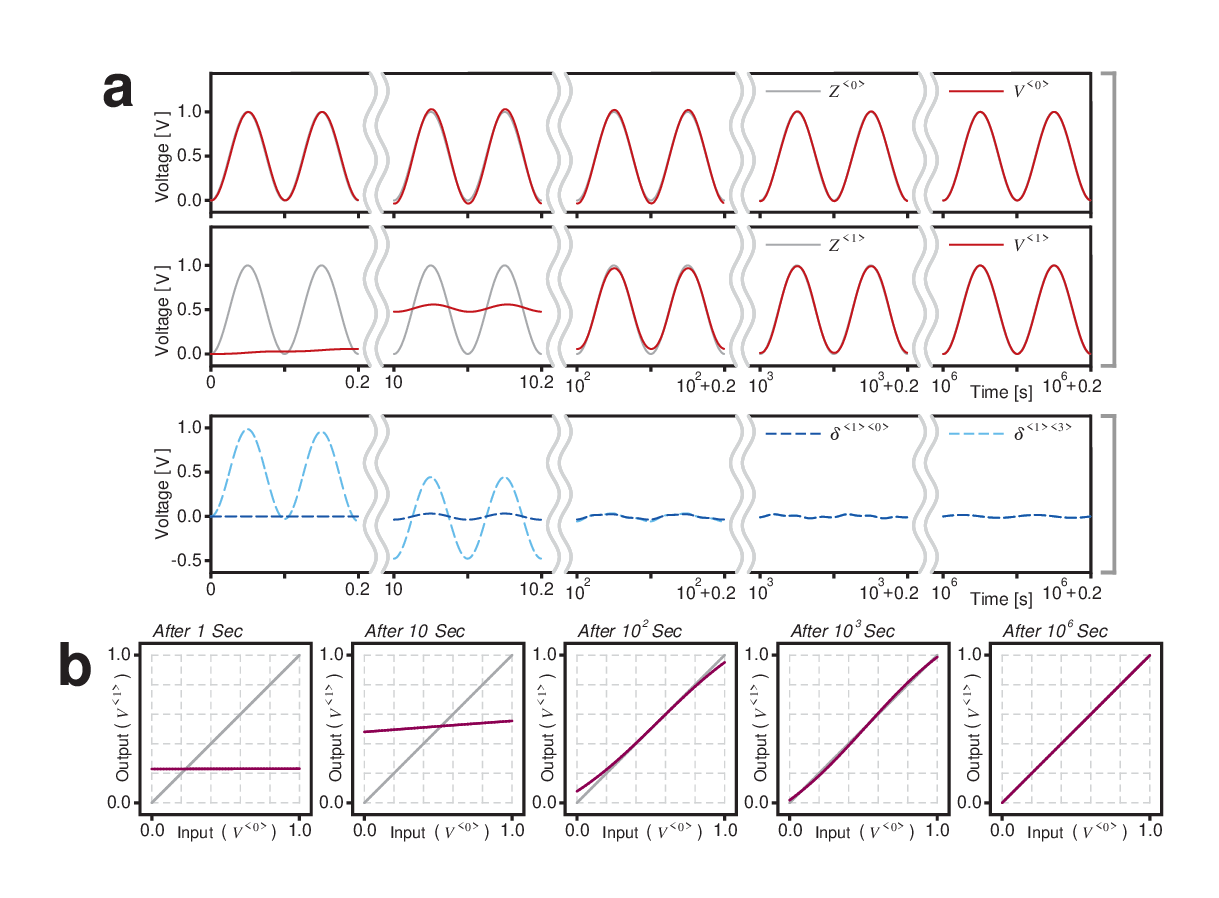}

   \vspace*{-3.0mm} 
    \caption{
      Simulation results for the Linear Case of the Learning Mode 
      in the one-dimensional model. 
      (a) Time-course of network parameters 
      when a set of sinusoidal waves 
      corresponding to the left drawings of Fig. 5 
      is applied to the input ports in the internal and external spaces. 
      (b) Transition of the input-output relationship 
      acquired in {\sl Internetwork}. 
      The graphs illustrate the results in which 
      {\sl Forward Subnet} detached from the model 
      is evaluated at five temporal stages. 
    } 

\end{figure}

Figure 4(a) illustrates the time-course of the network parameters.
Here, \( T = 1.0 ~{\sf S} \) and \( S = 1.0 ~{\sf S} \).
In simulation experiments, 
we observe the dynamics for \( 10^{6} \) seconds from the initial state 
and show the results of five intermediate stages in Fig. 4(a). 
The upper, middle, and lower graphs respectively denote 
the input \( Z^{<0>} \) and the output \( V^{<0>} \) in the internal space, 
the input \( Z^{<1>} \) and the output \( V^{<1>} \) in the external space, 
and the input signal to {\sl Backward Subnet} \( \delta^{<1><3>} \) and 
the output signal from {\sl Backward Subnet} \( \delta^{<1><0>} \). 
In the early stage, 
the figure clearly shows that 
the output signal from {\sl Backward Subnet} \( \delta^{<1><0>} \) 
almost equals zero. 
Therefore, in the internal space without any influence from the other factors, 
the output of the dynamical neuron \( V^{<0>} \) 
converges smoothly to the input value \( Z^{<0>} \) 
based on its time constant, 
and \( V^{<0>} \) and \( Z^{<0>} \) almost overlap. 
In the external space, on the other hand, 
the output of {\sl Forward Subnet} \( V^{<1>} \) is very small 
early on, 
and differs greatly from the input \( Z^{<1>} \). 
Thus, since the input to {\sl Backward Subnet} has a large value, 
learning in {\sl Internework} progresses adequately; 
the values of the synaptic connections in it are kept very small 
in the early stage of the time-course, 
so the output signal from {\sl Backward Subnet} remains small. 
This is why the output in the internal space \( V^{<0>} \) 
almost corresponds to the input from the outside \( Z^{<0>} \) 
as stated above. 
Learning in {\sl Internetwork} successfully advances with time; 
finally \( V^{<0>} \) goes to \( Z^{<0>} \), 
as well as \( V^{<1>} \) goes to \( Z^{<1>} \).

% Figure 05
\begin{figure}[!t]

    \hspace*{-6.5mm} 
    \includegraphics[scale=0.84]{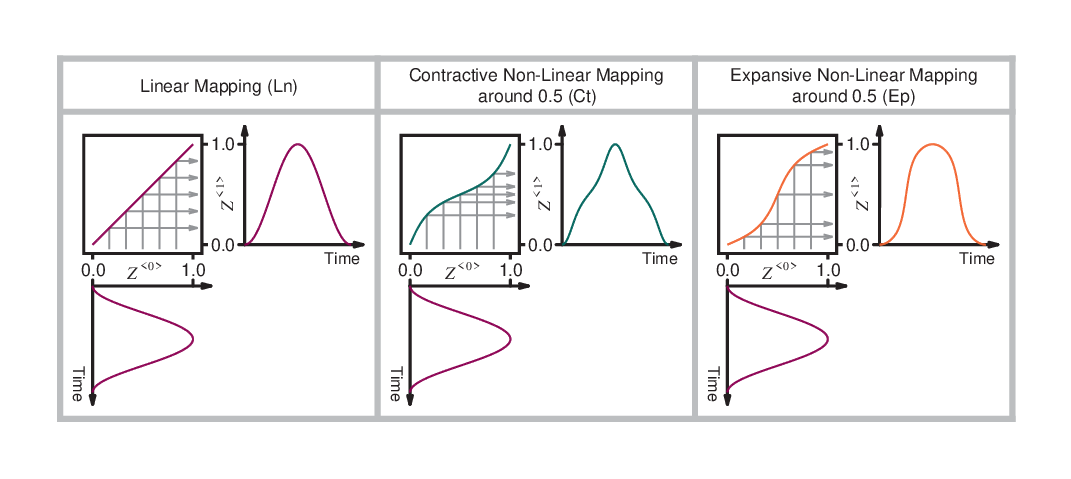}

    \vspace*{-4.0mm} 
    \caption{
      Three sets of sinusoidal and quasi-sinusoidal signals 
      supposed for testing network dynamics in the one-dimensional model. 
      An input signal with sinusoidal waveform for the internal space 
      is converted to that for the external space 
      employing three types of functions. 
    } 

\end{figure}

We can observe very interesting phenomena on the way to the final state, 
especially \( 10 \) or \( 10^{2} \) seconds from the beginning. 
At those times, the maximum amplitude of 
the output in the internal space \( V^{<0>} \) exceeds 
that of the input in the internal space \( Z^{<0>} \). 
Because the error signal from {\sl Backward Subnet} \( \delta^{<1><0>} \) 
is added to the dynamical neurons in the internal space, 
\( V^{<0>} \) increases to some extent. 
Since the architecture and the dynamics of our proposed network model 
are derived from the energy function, 
a steeper descent in the energy field is chosen 
in cooperation with the activity and learning dynamics. 
This collaborative effect produces a phenomenon in the graphs 
that contributes to the learning's acceleration and efficiency. 
We also investigated how {\sl Internetwork} can acquire 
a mapping relationship over time by isolating its part from the model. 
Figure 4(b) shows the input-output relationships of detached {\sl Forward Subnet} 
at five temporal stages. 
Although a rich input-output relationship is not attained in the early stage, 
a mapping relationship in {\sl Internetwork} gradually grows 
and becomes a complete linear one between \( 0.0 \) and \( 1.0 \).

\vspace{2mm} 
\noindent 
(2) Non-Linear Mapping

\vspace{2mm} 
\noindent 
In the previous simulation, we applied identical inputs 
in the internal and external spaces. 
It is quite interesting to study situations 
in which the waveforms of the input signals 
given to the internal and external spaces are different. 
As shown in Fig. 5, 
an input signal with a regular sinusoidal waveform for the internal space 
is converted to a sinusoidal or quasi-sinusoidal one 
for the external space using three types of functions.
In the middle and right cases, 
we employ, for non-linear conversion, a function in which 
a sine wave is superimposed to a straight line \( y = x \); 
the maximal rising or falling inclination 
of the superimposed sine wave is set here to \( 0.4 \) or \( -0.4 \). 

A regular sine wave used in the experiments for Figs. 4(a) and (b) 
corresponds to that on the left of Fig. 5, 
and the pre-converted and post-converted waves are identical 
since the conversion is due to a linear straight line through the origin; 
we call this the Linear (Mapping) Case. 
In contrast, the converting function in the middle case of Fig. 5 is 
expansive at the edges near \( 0.0 \) and \( 1.0 \)
but contractive around \( 0.5 \), 
so the finally produced periodic signal becomes an acute sinusoidal wave, 
called the Contractive (Mapping) Case. 
The converting function on the right of Fig. 5 is 
contractive at the edges near \( 0.0 \) and \( 1.0 \)
but expansive around \( 0.5 \), 
and the finally produced periodic signal 
leads to a sinusoidal wave rounded at the end, 
called the Expansive (Mapping) Case.

% Figure 06(a)(b)
\begin{figure}[!t]

    \hspace*{-6.0mm} 
    \includegraphics[scale=0.72]{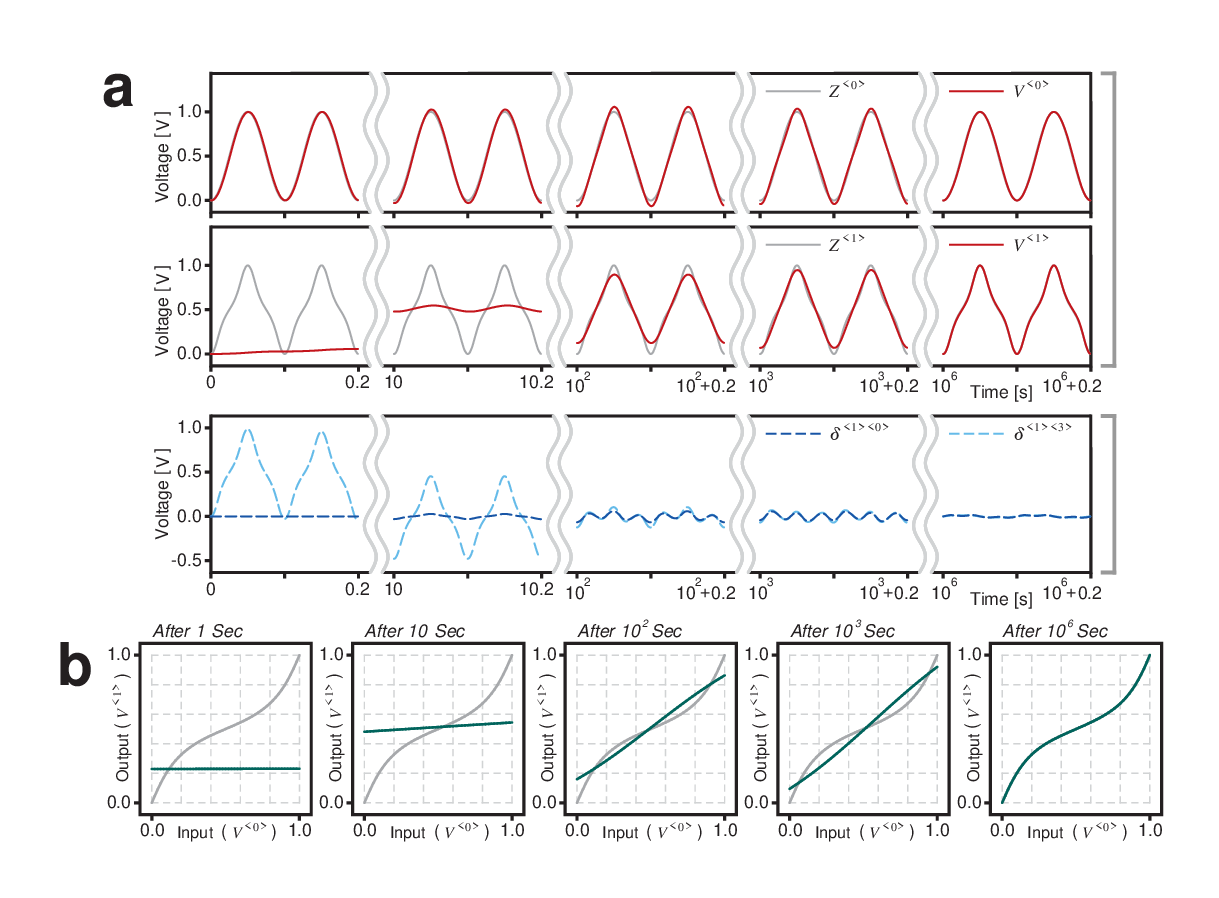}

    \vspace*{-3.0mm} 
    \caption{
      Simulation results for the Contractive Case of the Learning Mode 
      in the one-dimensional model. 
      (a) Time-course of network parameters 
      when a set of sinusoidal and quasi-sinusoidal waves 
      shown in the middle drawings of Fig. 5 
      is applied to the input ports in the internal and external spaces. 
      (b) Transition of the input-output relationship 
      acquired in {\sl Internetwork}. 
      The graphs illustrate the results in which 
      {\sl Forward Subnet} detached from the model 
      is evaluated at five temporal stages. 
    }

\end{figure}

% Figure 07(a)(b)
\begin{figure}[!t]

    \hspace*{-6.0mm} 
    \includegraphics[scale=0.72]{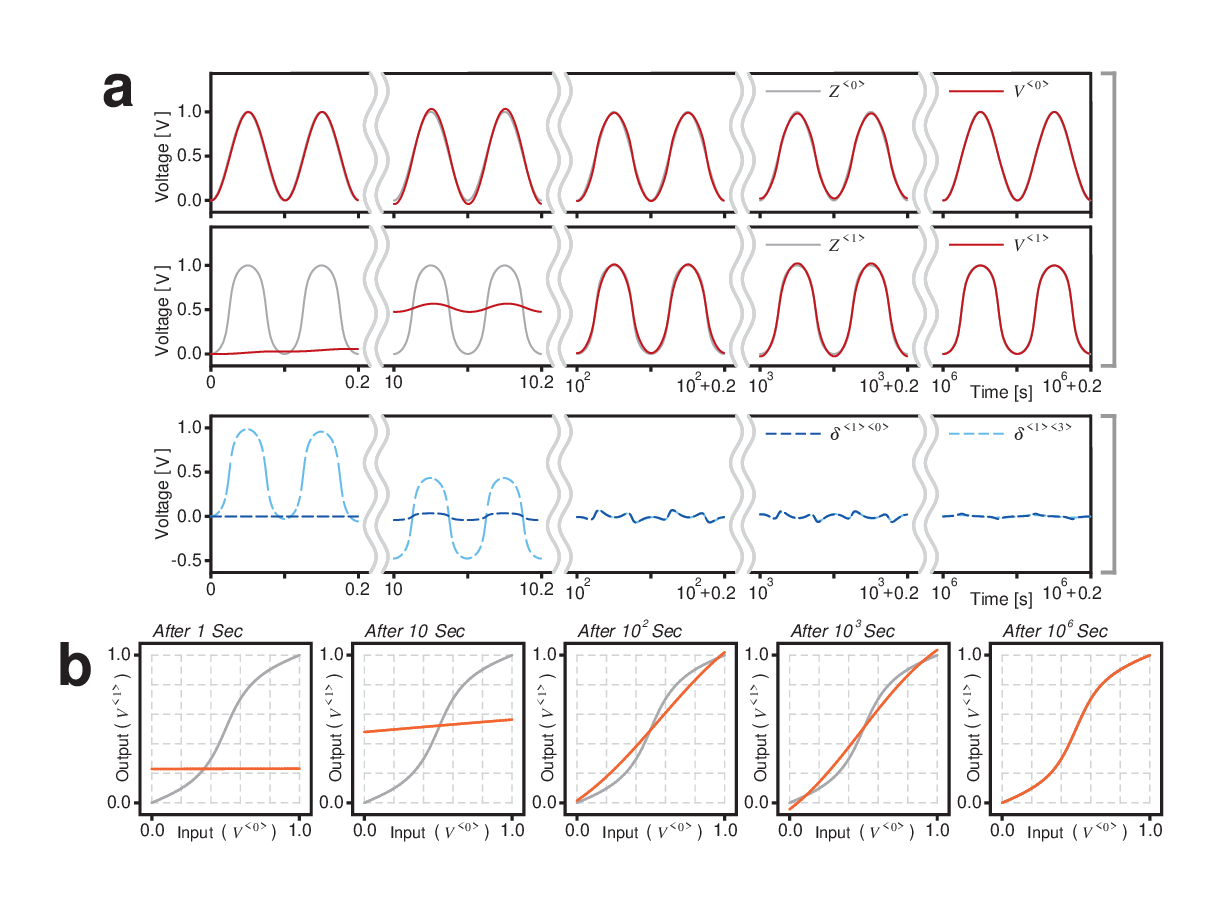}

    \vspace*{-3.0mm} 
    \caption{
      Simulation results for the Expansive Case of the Learning Mode 
      in the one-dimensional model. 
      (a) Time-course of network parameters 
      when a set of sinusoidal and quasi-sinusoidal waves 
      shown in the right drawings of Fig. 5 
      is applied to the input ports in the internal and external spaces. 
      (b) Transition of the input-output relationship 
      acquired in {\sl Internetwork}. 
      The graphs illustrate the results in which 
      {\sl Forward Subnet} detached from the model 
      is evaluated at five temporal stages. 
    } 

\end{figure}

Figure 6 is a simulation result 
in which we prepared a set of periodic waves in the middle case of Fig. 5, 
and applied the non-converted sinusoidal wave 
to the input port in the internal space 
and the converted one to the input port in the external space, respectively. 
Here, the frequency of both waves is set to \( 10.0 ~{\sf Hz} \). 
Figure 6(a) illustrates the time-course of the states in the model, 
and Fig. 6(b) shows the transition 
in the growth of the mapping relationship of {\sl Internetwork}; 
both graphs are depicted at five temporal stages 
as in the case of Figs. 4(a) and (b). 
Figure 7 shows the simulation result in which 
a set of periodic waves in the right case of Fig. 5 
is applied to the input ports in the internal and external spaces. 
The frequency of both input signals is also \( 10.0 ~{\sf Hz} \). 
Figure 7(a) shows the time-course of the states in the network, 
and Fig. 7(b) depicts the transition 
in the growth of the mapping relationship of {\sl Internetwork}.
The varing tendencies of the signals in Figs. 6(a) and (b) or Figs. 7(a) and (b) 
are basically common to those of Figs. 4(a) and (b), 
and the outputs \( V^{<0>} \) and \( V^{<1>} \) 
correspond respectively to the inputs \( Z^{<0>} \) and \( Z^{<1>} \) over time. 
Although the convergence times are different, 
the input-output relationships to be expected are eventually acquired; 
the mapping in the middle case of Fig. 5 is obtained in Fig. 6(b), 
and that in the right case of Fig. 5 is shown in Fig. 7(b). 

Scrutinizing the developmental process of these three cases, 
however, differences and interesting phenomena can be seen. 
In Figs. 4(a), 6(a), and 7(a), 
the output in the external space \( V^{<1>} \) gradually becomes larger. 
It does not always exceed 
the input in the external space \( Z^{<1>} \) in Figs. 4(a) and 6(a), 
although in Fig. 7(a), \( V^{<1>} \) slightly exceeds \( Z^{<1>} \) 
after \( 10^{3} \) seconds. 
It is also common to Figs. 4(a), 6(a), and 7(a) 
that, in the internal space after \( 10 \) seconds, 
the amplitude of \( V^{<0>} \) is already slightly larger 
than that of \( Z^{<0>} \). 
In Fig. 4(a) and Fig. 6(a), 
the amplitude of the output in the internal space \( V^{<0>} \) 
sometimes further increases and then decreases, 
but does not go below 
that of the input in the internal space \( Z^{<0>} \). 
On the contrary, in Fig. 7(a), 
the amplitude of \( V^{<0>} \) 
continues to gradually decrease after \( 10 \) seconds, 
and it nearly equals that of \( Z^{<0>} \) after \( 10^{2} \) seconds 
and is slightly smaller after \( 10^{3} \) seconds. 
Specifically, in Fig. 7(a) after \( 10^{3} \) seconds, 
the amplitude of the output \( V^{<0>} \) is a little smaller 
than that of the input \( Z^{<0>} \) in the internal space, 
and the amplitude of the output \( V^{<1>} \) is slightly larger 
than that of the input \( Z^{<1>} \) in the external space. 
This situation, that does not appear in 
Figs. 4(a) and 6(a) of the Linear and Contractive Cases, 
reflects the acquisition state of the mapping relationship in {\sl Internetwork}. 
Seeing Fig. 7(b) after \( 10^{3} \) seconds, 
the output of {\sl Forward Subnet} ranges from \(-0.044 \) to \( 1.035 \)
against the input to {\sl Forward Subnet} 
covering the interval from \( 0.0 \) to \( 1.0 \). 
Such a phenomenon, 
in which the output exceeds the input once during learning, 
can generally occur in a static multi-layered neural network 
without altering the range of input to the network, 
depending on the training set of the input-output relationship. 
However, of particular interest is the result 
in which the proposed network 
can successfully acquire the mapping relationship 
on the right of Fig. 5 (Expansive Case) in {\sl Internetwork}, 
varying the amplitude of an input signal to {\sl Internetwork} as mentioned above. 
Such behavior also seems to be caused by a synergistic effect 
due to the cooperation of the activity and learning dynamics. 

These observations show that, 
depending on the signal shape of the input from the outside, 
the mapping relationships acquired in {\sl Internetwork} 
as well as the learning speed and how synergistic effects appear 
are different. 
This might be caused 
by qualitative and quantitative differences 
between signals propagating through {\sl Backward Subnet}. 
We will further discuss this point in the following sections.

% 3.2 Learning Mode in Two-Dimensional Model 
\subsection{Learning Mode in Two-Dimensional Model}

\vspace{1mm} 
\noindent 
(1) Linear Mapping

\vspace{2mm} 
\noindent
In this subsection, we consider a two-dimensional model 
based on the results observed in a one-dimensional model. 
Since two input signals are necessary 
for each input port in the internal and external spaces 
in the two-dimensional model, 
we have to examine what kind of signals to give to the input ports. 
When we applied identical sinusoidal waves 
to the input ports in the internal and external spaces 
for the one-dimensional model, 
we obtained a linear mapping relationship 
that ranged from \( 0.0 \) to \( 1.0 \). 
If only one kind of sinusoidal wave is applied 
to four inputs (two inputs for two axes in the internal space 
and two inputs for two axes in the external space) 
for the two-dimensional model, 
instead of a mapping relationship with a two-dimensional extent, 
one degenerated merely into a single dimension may be acquired. 
In the following simulation, 
\( T_{ii} = 1.0 ~{\sf S} \) and \( S_{ii} = 1.0 ~{\sf S} \), 
~\( (i = 0, 1) \), 
and we set the other \( T \) and \( S \) values to \( 0 \). 
In a two-dimensional model, 
we employ sixteen hidden neurons in {\sl Internetwork}. 
The other conditions are basically common to those in the one-dimensional model. 

Figure 8 depicts how widely and densely two kinds of sinusoidal waves 
can cover a two-dimensional plane
when one is used to express a point on the horizontal axis 
and the other on the vertical axis. 
In terms of acquiring a two-dimensional mapping relation, 
evenly covering the entire region of the plane is better, 
and this mechanism is similar to that of the well-known Lissajous curves. 
Even in simple sinusoidal waves, 
we can arbitrarily arrange a combination of the frequency and the phase. 
In Fig. 8, we show an example in which 
six cases as frequency relationships, 
\( F_{1} = F_{0} \), \( F_{1} = (21/20) * F_{0} \), 
\( F_{1} = (20/19) * F_{0} \), \( F_{1} = 2 * F_{0} \), 
\( F_{1} = 5 * F_{0} \), and \( F_{1} = 20 * F_{0} \) 
from the upper to the bottom, 
and four cases as phase relationships, 
\( 0^{\circ} \), \( 90^{\circ} \), 
\( 180^{\circ} \), and \( 270^{\circ} \) 
from the left to the right are chosen. 
When the two frequencies are slightly different, e.g., as in the cases of 
\( F_{1} = (21/20) * F_{0} \) and \( F_{1} = (20/19) * F_{0} \), 
the entire region of the plane can evenly be covered. 
Even though the frequency difference between these two cases is small, 
each has its own characteristics. 
The coverage area is not influenced by any phase shifts 
in the former case, and it becomes denser 
in the latter case with a phase shift 
of \( 90^{\circ} \) or \( 270^{\circ} \) in particular. 
When \( F_{0} \) and \( F_{1} \) have a frequency relation of integral multiples, 
in contrast, the coverage area becomes wider and denser 
with the increase of its multiple ratio, 
although it is subject to the influence of the phase difference 
especially in the case of a small ratio of integral multiples. 
We chose three cases, 
\( F_{1} = (20/19) * F_{0} \), \( F_{1} = 20 * F_{0} \), 
and \( F_{1} = 5 * F_{0} \) without any phase shifts, 
and evaluated how their network states vary 
with the progress of the total dynamics.

% Figure 08
\begin{figure}[!t]

    \hspace*{-7.0mm} 
    \includegraphics[scale=0.80]{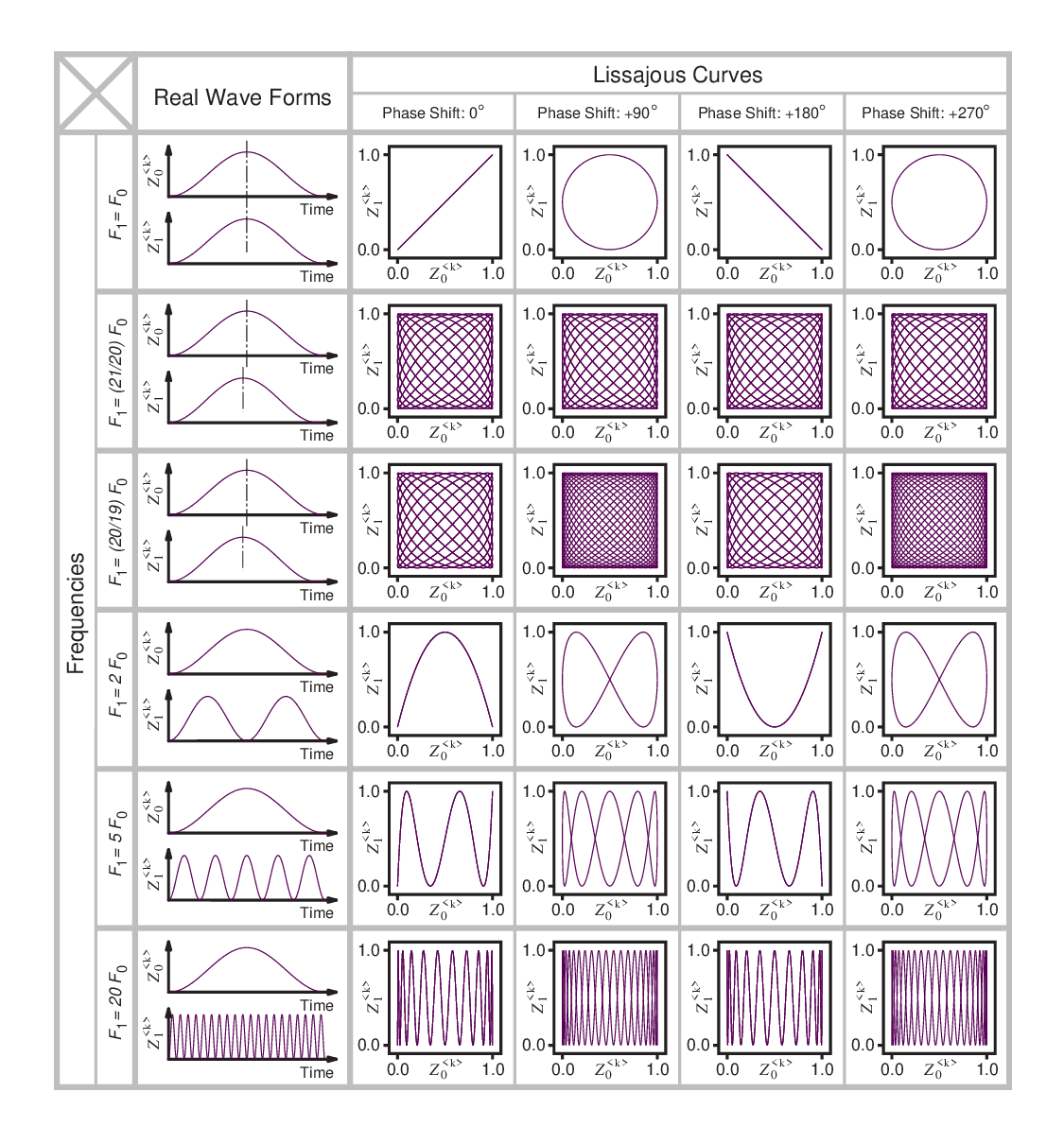}

    \vspace*{-3.0mm} 
    \caption{
      Relationship between two kinds of sinusoidal waves 
      with different frequencies and phases in the Linear Case. 
      The graphs show how widely and densely two signals 
      for horizontal and vertical axes 
      can cover a two-dimensional plane. 
      Note the similarity to Lissajous curves. 
    } 

\end{figure}

% Figure 09(a)(b)
\begin{figure}[!t]

    \hspace*{-7.0mm} 
    \includegraphics[scale=0.72]{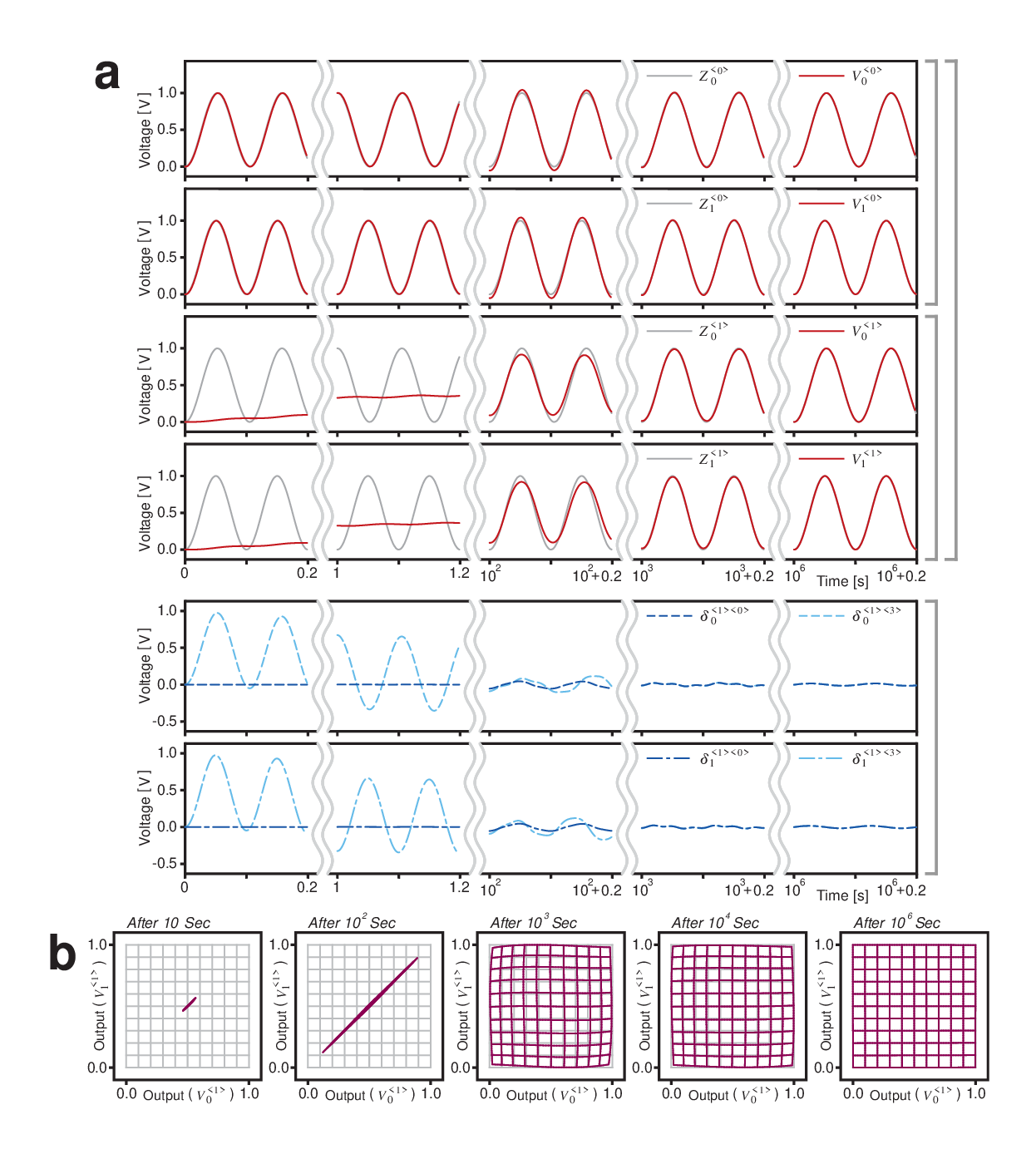}

    \vspace*{-3.0mm} 
    \caption{
      Simulation results for the Linear Case of the Learning Mode 
      in the two-dimensional model. 
      Frequency relationship between two sinusoidal waves 
      for horizontal and vertical axes is \( F_{1} = (20/19) * F_{0} \). 
      (a) Time-course of network parameters. 
      (b) Transition of the input-output relationship 
      acquired in {\sl Internetwork} (detached {\sl Forward Subnet}). 
    } 

\end{figure}

Figure 9 shows an example with \( F_{1} = (20/19) * F_{0} \), 
in which a non-distorted regular sinusoidal wave with \( 9.5 ~{\sf Hz} \) 
is applied to \( Z^{<0>}_{0} \) and \( Z^{<1>}_{0} \) for the horizontal axes, 
and that with \( 10.0 ~{\sf Hz} \) 
is given to \( Z^{<0>}_{1} \) and \( Z^{<1>}_{1} \) for the vertical axes. 
Figure 9(a) illustrates the time-course at the signal level, 
and Fig. 9(b) depicts the progress of the mapping relationship 
acquired in {\sl Internetwork} on the total dynamics. 
In Fig. 9(a), the phases of \( Z^{<0>}_{0} \) and \( Z^{<1>}_{0} \) 
from \( 1.0 ~{\sf s} \) to \( 1.2 ~{\sf s}\) 
are roughly opposite of those from \( 0.0 ~{\sf s} \) to \( 0.2 ~{\sf s} \); 
this is because one period of the signal with \( 9.5 ~{\sf Hz} \) 
is \( 2/19 \) seconds. 
In the two-dimensional model, 
both activity and learning dynamics 
progress cooperatively and smoothly through such a stage that 
the amplitudes of \( V^{<0>}_{0} \) and \( V^{<0>}_{1} \) 
exceed those of \( Z^{<0>}_{0} \) and \( Z^{<0>}_{1} \) 
in the same way as the one-dimensional model. 
In the end, the outputs in the external space 
 \( V^{<1>}_{0} \) and \( V^{<1>}_{1} \) 
almost coincide with the inputs in the external space 
 \( Z^{<1>}_{0} \) and \( Z^{<1>}_{1} \). 
From the viewpoint of the acquisition 
of the mapping relationship in {\sl Internetwork}, 
the progress over time is very smooth as shown in Fig. 9(b). 
{\sl Internetwork} finally acquires a two-dimensional linear mapping 
within a range strictly from \( 0.0 \) to \( 1.0 \). 
When we concentrate on the state after \( 10^{2} \) seconds in Fig. 9(a), 
the learning of {\sl Internetwork} appears to be already well-advanced, 
judging from the situation in which 
the amplitudes of \( V^{<1>}_{0} \) and \( V^{<1>}_{1} \) become fairly large. 
However, this is not the case. 
By carefully looking at the signals \( V^{<1>}_{0} \) and \( V^{<1>}_{1} \), 
\( V^{<1>}_{0} \) moves to the left of \( Z^{<1>}_{0} \) 
and seems to be going to a higher frequency, 
and in contrast, 
\( V^{<1>}_{1} \) goes to the right of \( Z^{<1>}_{1} \) 
and appears to be moving to a lower frequency. 
This behavior creates an effect in which 
the frequencies of two input signals 
applied for the horizontal and vertical axes 
are getting closer to each other, 
although the intervals between zero-crossings remain constant 
and those frequencies do not actually change at all. 
Checking the state after \( 10^{2} \) seconds in Fig. 9(b), 
the mapping relationship obtained at this moment is not two-dimensional; 
it remains almost on a one-dimensional straight line. 
This phenomenon, in which such an interesting way of learning is chosen, 
also reflects the mutual interaction of the activity and learning dynamics 
derived from the energy function through its minimizing procedure.

\vspace{2mm} 
\noindent 
(2) Non-Linear Mapping 

\vspace{2mm} 
\noindent 
Assuming quasi-sinusoidal waves in the middle panel of Fig. 5 (the Contractive Case), 
Fig. 10 shows, in a way similar to Fig. 8, how widely and densely 
those signals with various relations between frequencies and phases 
can cover a two-dimensional plane. 
The relations between frequencies and phases 
basically resemble those in Fig. 8. 
However, even if the relation between frequencies is the same 
as the case with non-distorted sinusoidal waves, 
some areas become denser and others are sparser 
depending on the degree of non-linearity for a converting function.

% Figure 10
\begin{figure}[!t]

    \hspace*{-7.0mm}
    \includegraphics[scale=0.80]{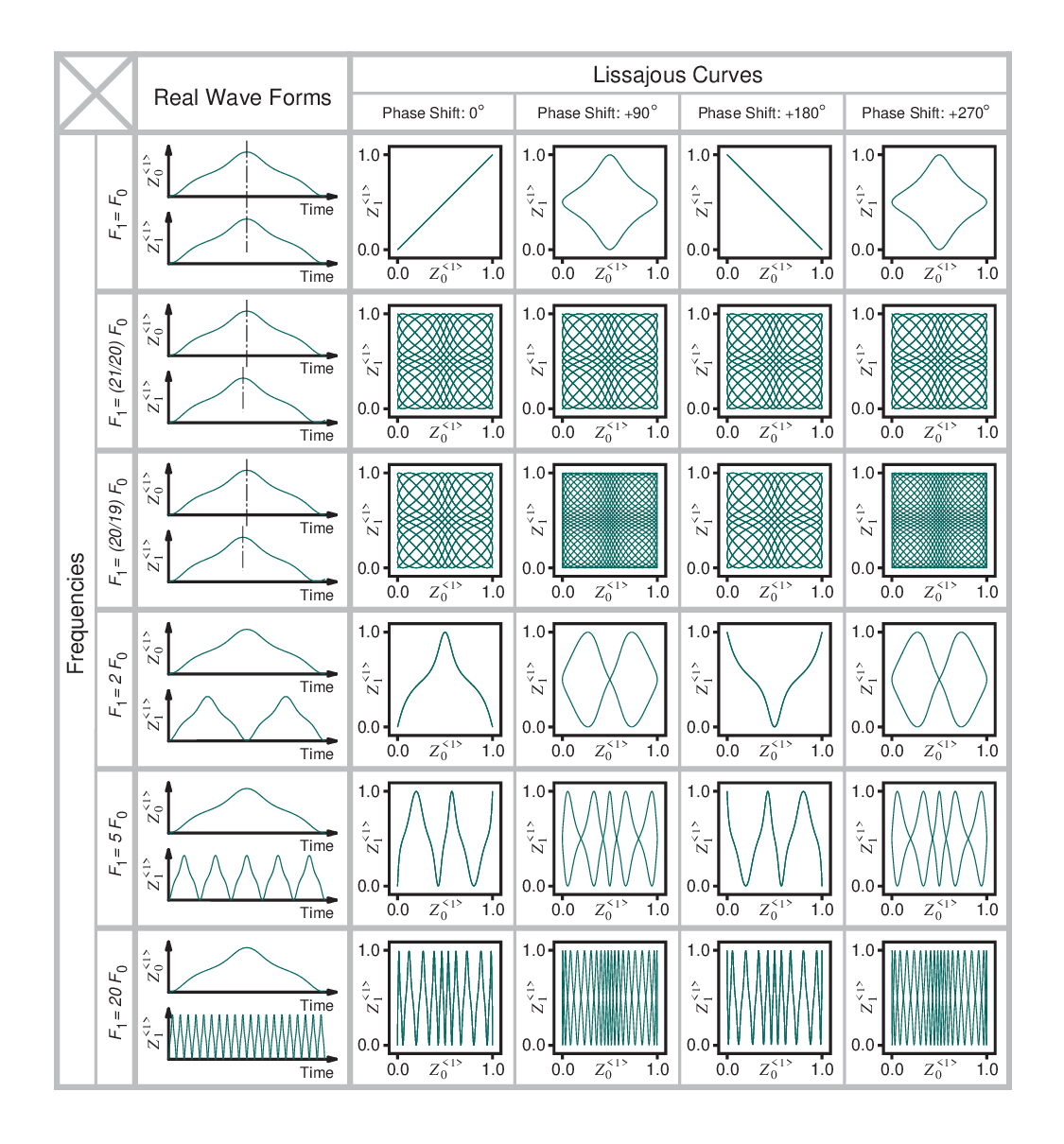}

    \vspace*{-3.0mm}
    \caption{
      Relationship between two kinds of quasi-sinusoidal waves 
      with different frequencies and phases in the Contractive Case. 
      The graphs show how widely and densely two signals 
      for horizontal and vertical axes 
      can cover a two-dimensional plane. 
      Also note the similarity to Lissajous curves. 
    } 

\end{figure}

% Figure 11(a)(b)
\begin{figure}[!t]

    \hspace*{-7.0mm} 
    \includegraphics[scale=0.72]{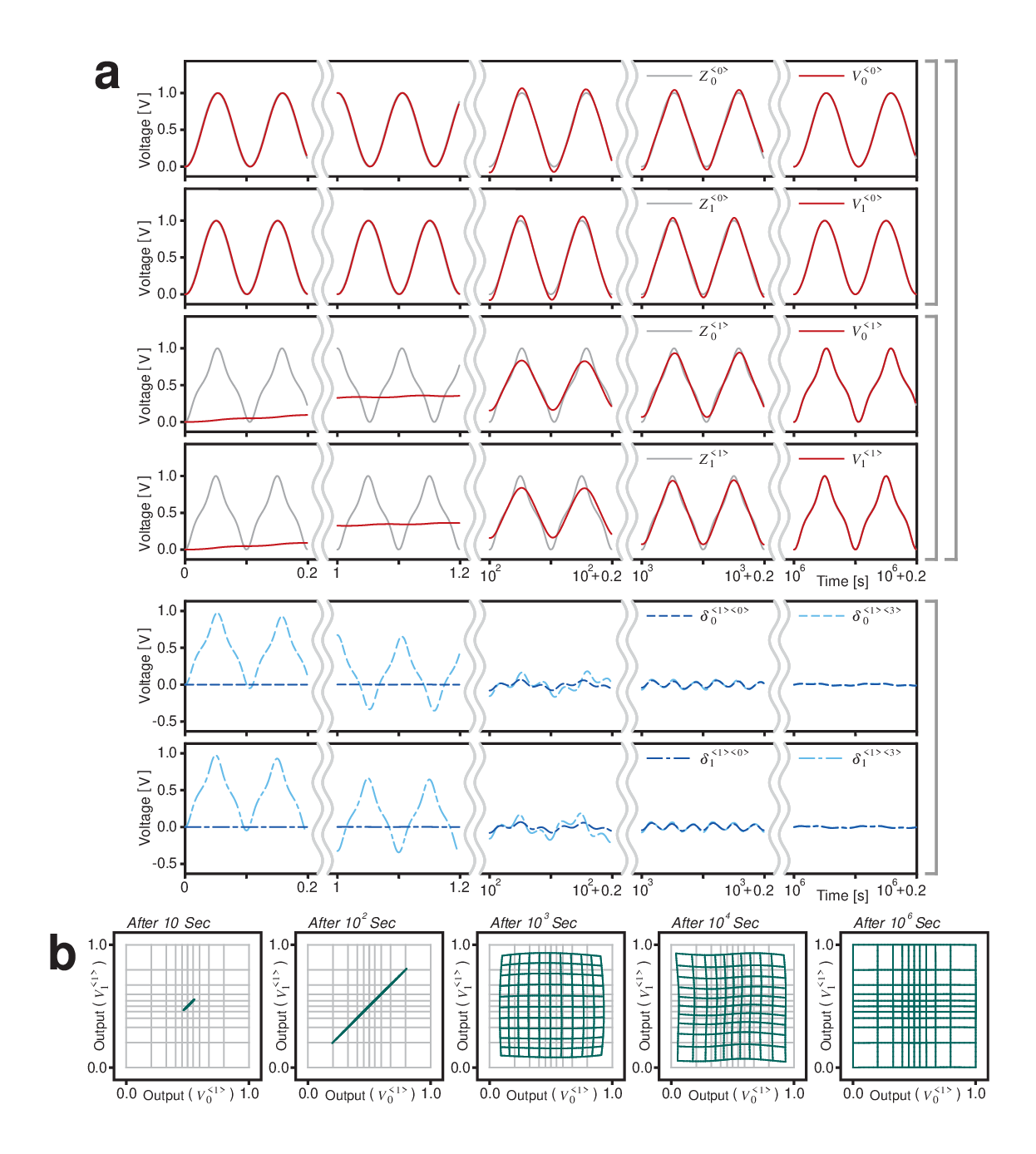}

    \vspace*{-3.0mm} 
    \caption{
      Simulation results for the Contractive Case of the Learning Mode 
      in the two-dimensional model. 
      Frequency relationship between two quasi-sinusoidal waves 
      for horizontal and vertical axes is \( F_{1} = (20/19) * F_{0} \). 
      (a) Time-course of network parameters. 
      (b) Transition of the input-output relationship 
      acquired in {\sl Internetwork} (detached {\sl Forward Subnet}). 
    } 

\end{figure}

Figure 11 shows a simulation result with \( F_{1} = (20/19) * F_{0} \) 
under no phase shift, in which 
the non-distorted regular sinusoidal waves 
with \( 9.5 ~{\sf Hz} \) for the horizontal axis 
and \( 10.0 ~{\sf Hz} \) for the vertical axis 
are respectively applied to \( Z^{<0>}_{0} \) and \( Z^{<0>}_{1} \), 
and the distorted sinusoidal waves of the Contractive Case 
with \( 9.5 ~{\sf Hz} \) for the horizontal axis 
and \( 10.0 ~{\sf Hz} \) for the vertical axis  
are respectively put to \( Z^{<1>}_{0} \) and \( Z^{<1>}_{1} \). 
Figure 11(a) shows the time-course of the dynamics at the signal level, 
and Fig. 11(b) shows the transition of the mapping relationship 
in {\sl Internetwork} over time. 
As is clear in Figs. 11(a) and (b), 
a mapping relationship that is contractive around \( 0.5 \) and 
expansive at the edges near \( 0.0 \) and \( 1.0 \) 
was successfully acquired through such a stage that 
the amplitudes of \( V^{<0>}_{0} \) and \( V^{<0>}_{1} \) 
exceed those of \( Z^{<0>}_{0} \) and \( Z^{<0>}_{1} \) 
after \( 10^{2} \) seconds. 
As for the mapping of the Contractive Case, 
the learning speed tends to be 
% 
%%%%% ----- Updated Part No. 01 (Begin) ----- 
%%%%% v1 
% slower 
%%%%% v2 
lower 
%%%%% ----- Updated Part No. 01 (End) ----- 
% 
in the same way as in the one-dimensional model, 
and this is quite obvious from a comparison of the states 
after \( 10^{3} \) seconds from the beginning in Figs. 9(b) and 11(b). 

When quasi-sinusoidal waves in the right panel 
of Fig. 5 (the Expansive Case) are employed, 
Fig. 12 depicts how widely and densely they can cover a two-dimensional plane. 
In a way similar to Fig. 10, 
some areas are denser but others are sparser
for the relation between lines.
Note, however, 
that the area where information necessary for learning is densely applied
changes depending on the quantitative differences of the non-linearity; 
the relations between the denser and sparser areas 
are reversed in Figs. 10 and 12, 
on the basis of Fig. 8 without any distortion of waves.

% Figure 12
\begin{figure}[!t]

    \hspace*{-7.0mm} 
    \includegraphics[scale=0.80]{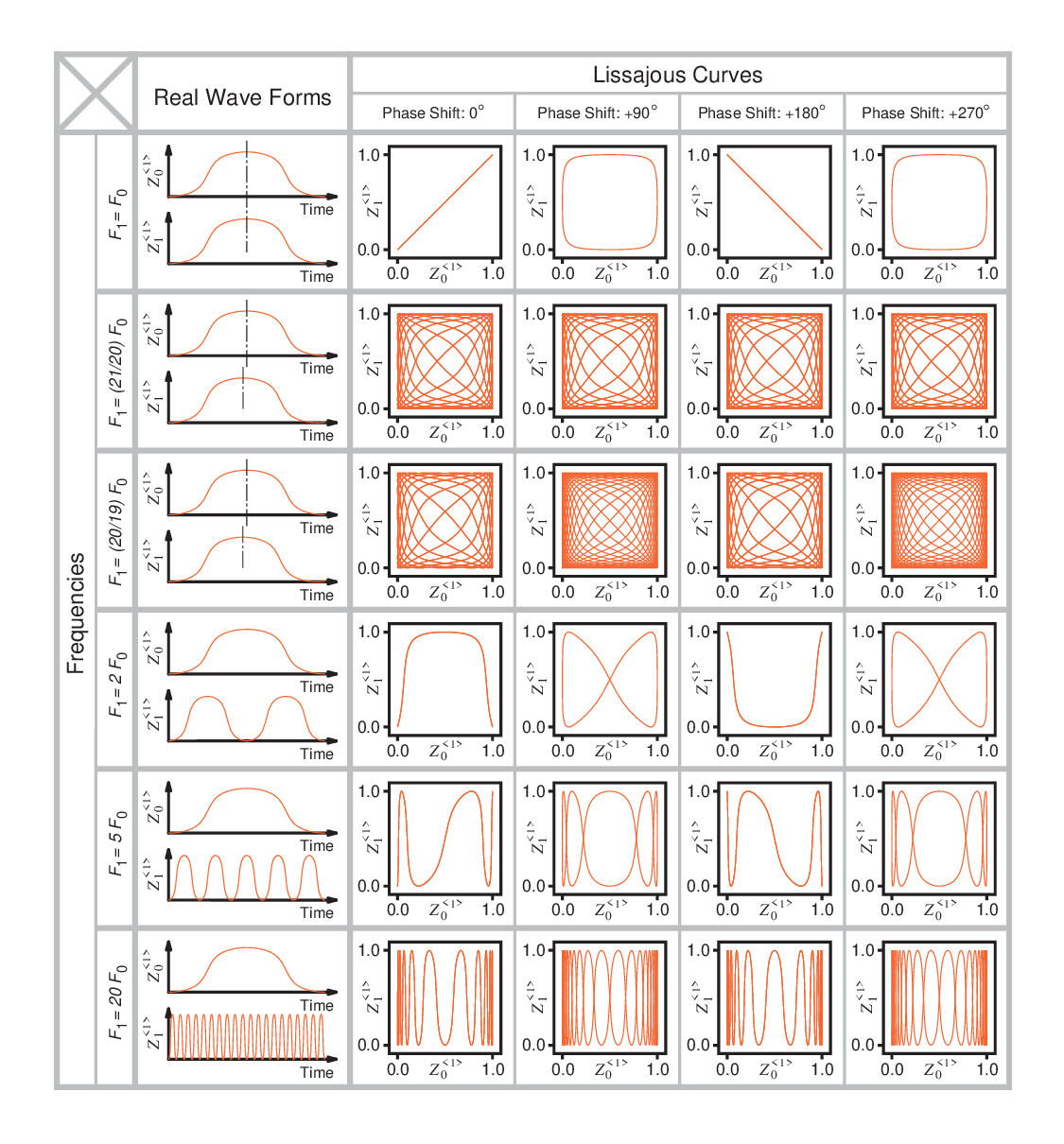}

    \vspace*{-3.0mm} 
    \caption{ 
      Relationship between two kinds of quasi-sinusoidal waves 
      with different frequencies and phases in the Expansive Case. 
      The graphs show how widely and densely two signals 
      for horizontal and vertical axes 
      can cover a two-dimensional plane. 
      Also note the similarity to Lissajous curves. 
    } 

\end{figure}

Figure 13 shows a simulation result with \( F_{1} = 20 * F_{0} \) under no phase shift, 
in which the relationship between frequencies \( F_{0} \) and \( F_{1} \) 
is quite unlike the above examples in the two-dimensional model. 
The non-distorted regular sinusoidal waves 
with \( 0.5 ~{\sf Hz} \) for the horizontal axis 
and \( 10.0 ~{\sf Hz} \) for the vertical axis 
are respectively applied to \( Z^{<0>}_{0} \) and \( Z^{<0>}_{1} \), 
and the distorted sinusoidal waves in the Expansive Case 
with \( 0.5 ~{\sf Hz} \) for the horizontal axis 
and \( 10.0 ~{\sf Hz} \) for the vertical axis  
are respectively put to \( Z^{<1>}_{0} \) and \( Z^{<1>}_{1} \). 
Figure 13(a) is the time-course of the dynamics at the signal level.
Comparing Fig. 13(a) with Fig. 9(a) or Fig. 11(a), 
the frequency of \( F_{1} \) in Fig. 13(a) seems extremely high at a glance, 
but \( F_{1} \) is commonly set to \( 10.0 {\sf Hz} \), and 
note that this impression is caused by the lower value setting of \( F_{0} \). 
Figure 13(b) shows the transition of the mapping relationships 
acquired in {\sl Internetwork} over time. 
As is clear from this result, also in the case of \( F_{1} = 20 * F_{0} \), 
a complete mapping relationship that is expansive around \( 0.5 \) 
and contractive at the edges near \( 0.0 \) and \( 1.0 \) 
can finally be acquired in {\sl Internetwork}.

% Figure 13(a)(b)
\begin{figure}[!t]

    \hspace*{-7.0mm} 
    \includegraphics[scale=0.72]{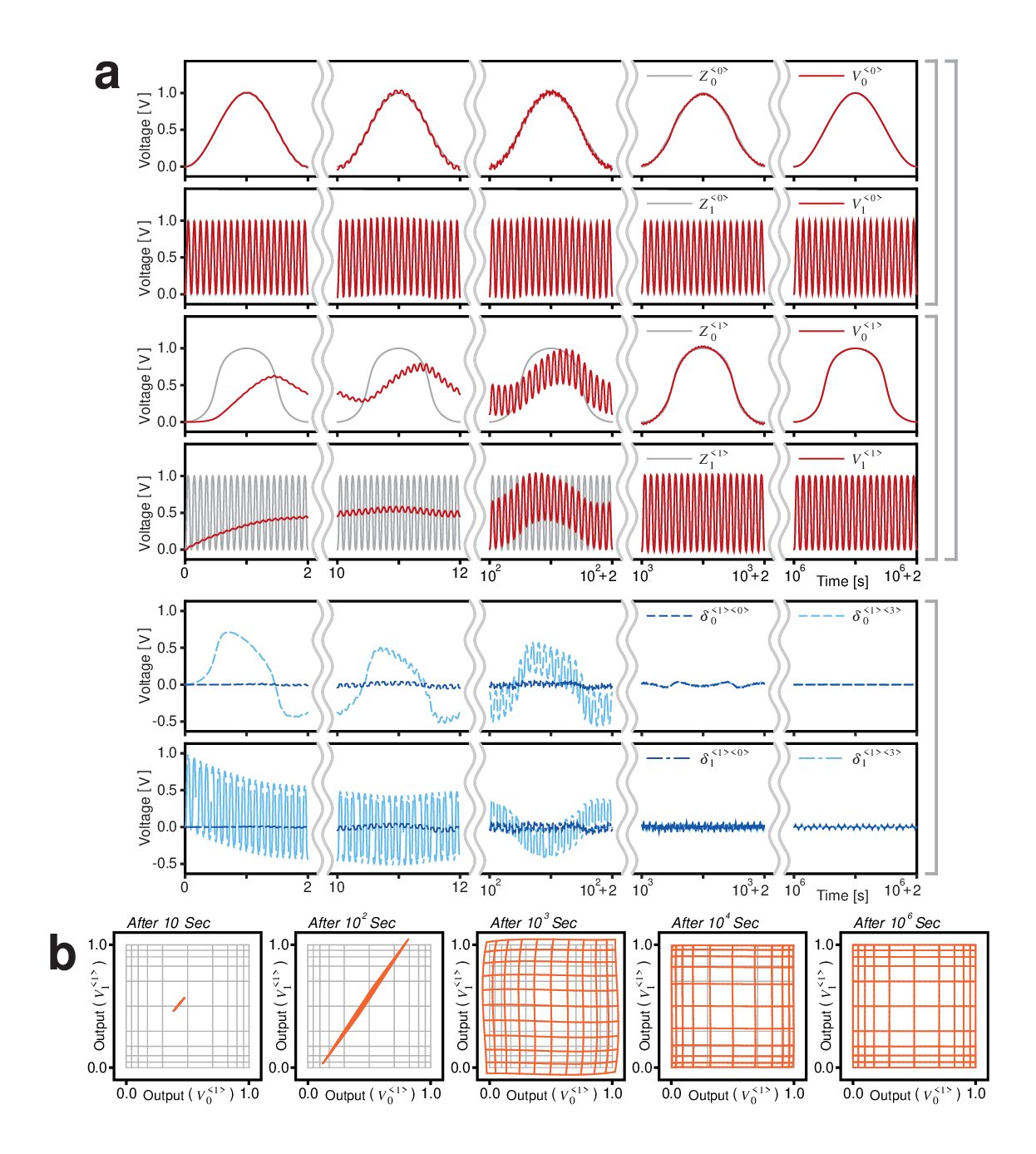}

    \vspace*{-3.0mm} 
    \caption{
      Simulation results for the Expansive Case of the Learning Mode 
      in the two-dimensional model. 
      Frequency relationship between two quasi-sinusoidal waves 
      for horizontal and vertical axes is \( F_{1} = 20 * F_{0} \). 
      (a) Time-course of network parameters. 
      (b) Transition of the input-output relationship 
      acquired in {\sl Internetwork} (detached {\sl Forward Subnet}). 
    } 

\end{figure}

Thus, we conclude that a two-dimensional mapping relationship 
between the internal and external spaces can successfully be obtained 
depending on the degree of the non-linear warping for periodic signals 
applied to the input ports in the internal and external spaces, 
when \( F_{0} \) and \( F_{1} \) have such a relation 
that either they are slightly different 
(Fig. 8 and Figs. 9(a) and (b) or Fig. 10 and Figs. 11(a) and (b)) 
or very different (Fig. 12 and Figs. 13(a) and (b)). 

When the frequency relationship between \( F_{0} \) and \( F_{1} \) 
takes a low integral multiple, such as 
\( F_{1} = 2 * F_{0} \) or \( F_{1} = 5 * F_{0} \) in Fig. 12, 
how do the network dynamics progress over time 
and what kind of results can be obtained?
Figure 14 shows a simulation result 
with \( F_{1} = 5 * F_{0} \) under no phase shift. 
Non-distorted regular sinusoidal waves 
with \( 2.0 ~{\sf Hz} \) for the horizontal axis 
and \( 10.0 ~{\sf Hz} \) for the vertical axis 
are respectively applied to \( Z^{<0>}_{0} \) and \( Z^{<0>}_{1} \), 
and the distorted sinusoidal waves in the Expansive Case 
with \( 2.0 ~{\sf Hz} \) for the horizontal axis 
and \( 10.0 ~{\sf Hz} \) for the vertical axis  
are respectively put to \( Z^{<1>}_{0} \) and \( Z^{<1>}_{1} \). 
Figure 14(a) indicates the time-course of the dynamics at the signal level, and 
Fig. 14(b) shows the transition of the mapping relationship 
in {\sl Internetwork} over time.
As is clear in Fig. 14(a), 
even when \( F_{0} \) and \( F_{1} \) are related with a small integral multiple, 
the outputs in the external space \( V^{<1>}_{0} \) and \( V^{<1>}_{1} \) 
eventually overlap with 
the inputs in the external space \( Z^{<1>}_{0} \) and \( Z^{<1>}_{1} \). 
As shown in Fig. 14(b), however, 
a complete two-dimensional mapping relation is not obtained, 
compared with the \( F_{1} = 20 * F_{0} \) case illustrated in Fig. 13(b). 
Note the \( F_{1} = 5 * F_{0} \) case with no phase shift in Fig. 12 again.
For this frequency relation, the information effective for learning 
is sparse within the square two-dimensional plane, and 
its situation seems to cause the above phenomenon. 
It is not hard to imagine that the result worsens when \( F_{1} = 2 * F_{0} \).
Even for this \( F_{1} = 5 * F_{0} \), 
the simulation results changes 
depending on the number and the gain of the hidden units in {\sl Internetwork}. 
In any cases, when \( F_{0} \) and \( F_{1} \) 
are related with a small integral multiple, 
the activity and learning dynamics themselves progress smoothly to a final state, 
but the mapping relationship acquired in {\sl Internetwork} does not become rich 
with all the generalization capability of a layered neural network.

% Figure 14(a)(b)
\begin{figure}[!t]

    \hspace*{-7.0mm} 
    \includegraphics[scale=0.72]{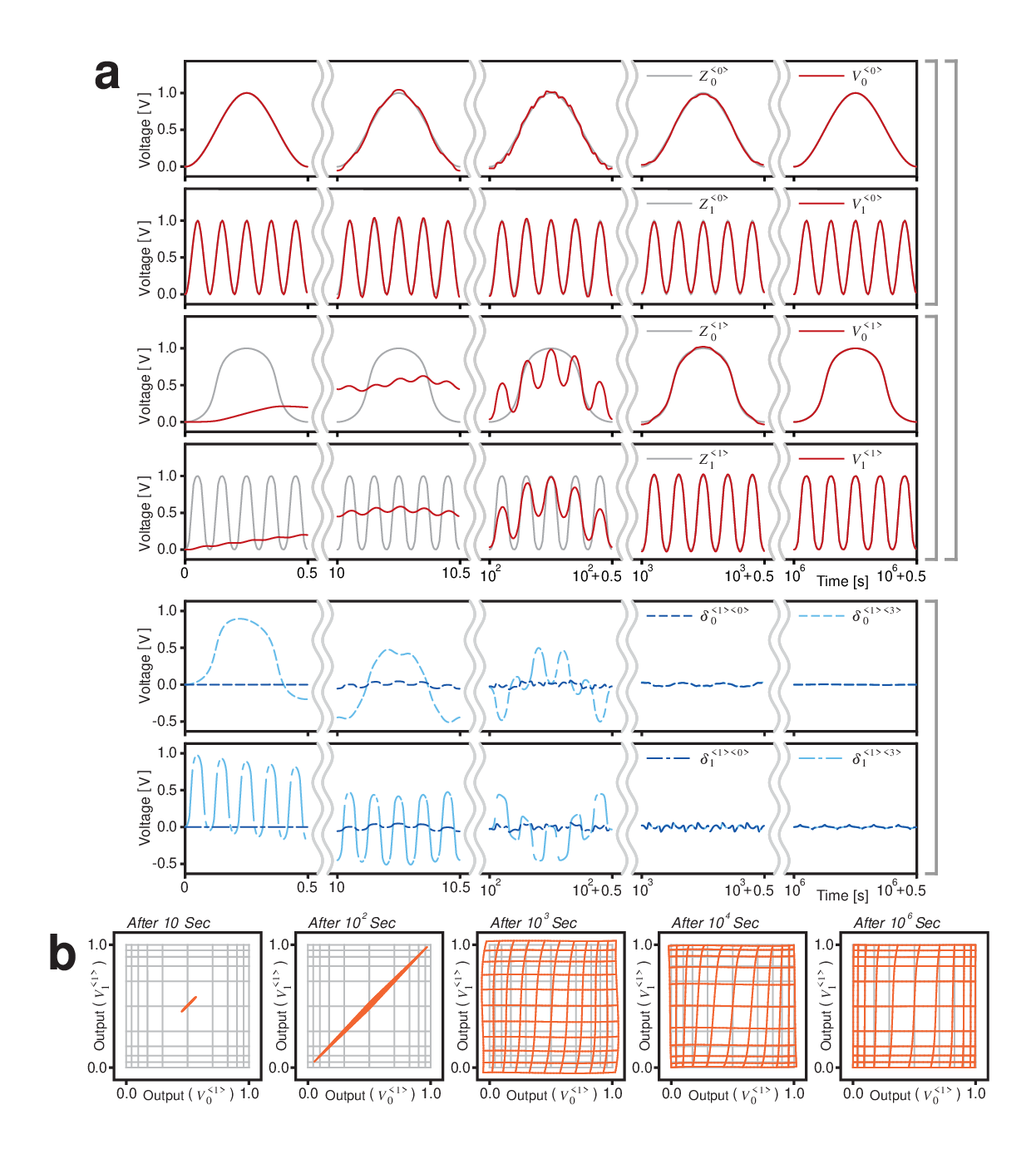}

    \vspace*{-5.0mm} 
    \caption{ 
      Simulation results for the Expansive Case of the Learning Mode 
      in the two-dimensional model. 
      Frequency relationship between two quasi-sinusoidal waves 
      for horizontal and vertical axes is \( F_{1} = 5 * F_{0} \). 
      (a) Time-course of network parameters. 
      (b) Transition of the input-output relationship 
      acquired in {\sl Internetwork} (detached {\sl Forward Subnet}). 
    } 

\end{figure}

% 4 Association Mode
\section{Association Mode}

% 4.1 Association Mode in One-Dimensional Model 
\subsection{Association Mode in One-Dimensional Model}

\noindent
In our model, as stated in Subsection 2.3, 
we distinguish two modes, Learning and Association, 
that depend on whether pairs (or just one pair) of 
a feedback path and an input port exist(s) 
in the internal and/or external space(s). 
In this section, we describe the dynamical behavior of the Association Mode 
in which Eq. (29) is supposed to be OFF 
and all other conditions are shared with those in Section 3. 
Let us consider the one-dimensional model first. 

Figure 15 shows the time-course of the network states from four initial points, 
when a pair of a feedback path and an input port 
exists only in the internal space (Type \#0 in Fig. 3) and 
a fixed value of \( 0.5 \) is applied as an input for the internal space. 
The obtained results are shown as a table with three rows. 
The upper row is the Contractive Case with a mapping relationship 
that is obtained by the simulation shown in Fig. 6, 
the middle one is the Linear Case based on the simulation shown in Fig. 4,
and the lower one is the Expansive Case based on the simulation shown in Fig. 7.
For the horizontal direction, 
each of the leftmost graphs shows the time-course of the output in the internal space, 
its right one is the mapping relationship already acquired in {\sl Internetwork}, 
the next right one shows the time-course of the output in the external space, 
and the rightmost one depicts the time-course 
of the output from {\sl Backward Subnet}. 
To focus on the initial responses, 
we only plot the variations for \( 4 ~{\sf ms} \) from the beginning.
Also to help grasp the meaning of each graph, 
we draw a block diagram of the whole network 
with gray wide arrow lines in the upper part 
and append an index to each graph. 
As is clear from {\tt 15-Ct-DL}, {\tt 15-Ln-DL}, and {\tt 15-Ep-DL},
in this Association Mode with a pair of a feedback path and an input port 
only in the internal space, 
no signal propagates through {\sl Backward Subnet} in principle.
So the output in the internal space \( V^{<0>} \) 
is simply converted in a static manner 
to the output in the external space \( V^{<1>} \) by {\sl Forward Subnet}. 
Therefore, in the Contractive Case in the upper row, for instance, 
external {\tt 15-Ct-V<1>} is entirely contractive 
against internal {\tt 15-Ct-V<0>}, and 
in the Expansive Case in the lower row, 
external {\tt 15-Ep-V<1>} is expansive overall to internal {\tt 15-Ep-V<0>}. 
In the Linear Case in the middle row, on the contrary, 
the internal {\tt 15-Ln-V<0>} and external {\tt 15-Ln-V<1>} graphs 
are completely the same.

% Figure 15
\begin{figure}[!t]

    \hspace*{-6.5mm} 
    \includegraphics[scale=0.86]{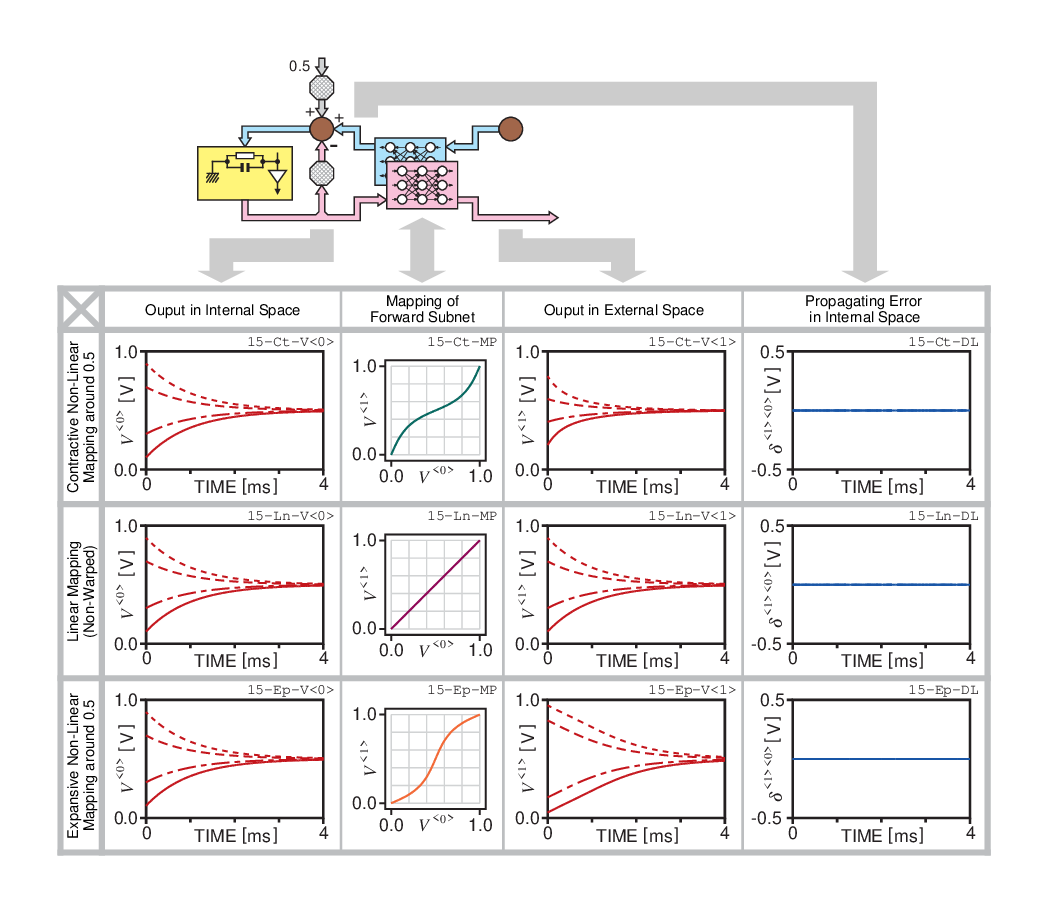}

    \vspace*{-3.0mm} 
    \caption{ 
      Simulation results for Type \#0 
      of the Association Mode in the one-dimensional model. 
      In this experiment, the place with a pair of a feedback path and an input port 
      is limited to the internal space. 
      The graphs include the Contractive, Linear, and Expansive Cases. 
    }

\end{figure}

Next, we consider the network behavior 
when a pair of a feedback path and an input port exists only in the external space. 
Notice the signal propagating through {\sl Backward Subnet} in this case. 
Figure 16 illustrates what kind of signal flows through {\sl Backward Subnet} 
for three mapping cases of {\sl Internetwork}. 
The lower right figure shows the linkage between 
a block diagram of the corresponding part and signal flows, 
and the lower left graph is the relation between 
the signals \textcircled{\scriptsize 1} and \textcircled{\scriptsize 2}. 
The upper five graphs depict the relation between 
the output from {\sl Forward Subnet} \textcircled{\scriptsize 2} 
and that from {\sl Backward Subnet} \textcircled{\scriptsize 4}, 
choosing five values as the input from the outside \textcircled{\scriptsize 3}. 
Since the upper five graphs and the lower left figure are related 
with signal \textcircled{\scriptsize 2}, 
we supplement their relation using multiple dashed arrow lines 
with different gray levels and widths. 
In the lower left and upper five graphs, 
green means the Contractive Case, 
orange corresponds to the Expansive Case, 
and purple indicates the Linear Case.

% Figure 16
\begin{figure}[!t]

    \hspace*{-6.0mm} 
    \includegraphics[scale=0.78]{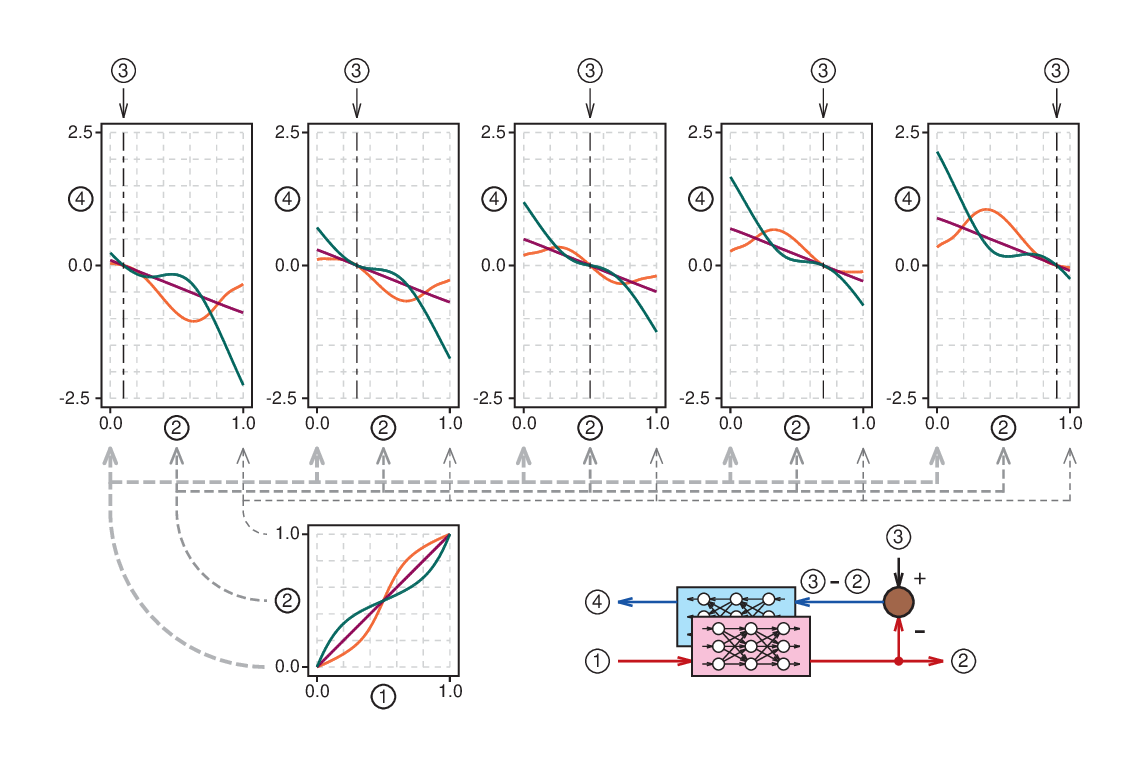}

    \vspace*{-3.0mm} 
    \caption{ 
      A conceptual diagram and graphs describing 
      signal flows through {\sl Backward Subnet} in Type \#1 of the Association Mode. 
      The lower right drawings show the relation 
      between a block diagram of the corresponding part and signal flows. 
      The lower left graph indicates the input-output relation in {\sl Forward Subnet}. 
      The upper five graphs show the relation between 
      the output from {\sl Forward Subnet} and that from {\sl Backward Subnet}, 
      choosing five values as the input from the outside. 
      In the lower left graph and the upper five graphs, 
      green, purple, and orange correspond 
      to the Contractive, Linear, and Expansive Cases. 
    } 

\end{figure}

The output in the internal space \textcircled{\scriptsize 1} 
is converted to the output in the external space \textcircled{\scriptsize 2} 
by {\sl Forward Subnet}. 
Computing the subtraction of \textcircled{\scriptsize 2} 
from \textcircled{\scriptsize 3}, 
\textcircled{\scriptsize 3} - \textcircled{\scriptsize 2} 
is applied to {\sl Backward Subnet}, 
where, as is clear from Eq. (13) in general, 
the product-sum operation of 
\( \partial V^{<1>}_{i^{\prime}} / \partial V^{<0>}_{i} \) and 
( \textcircled{\scriptsize 3} - \textcircled{\scriptsize 2} ) 
for \( i^{\prime} \) is executed. 
Since the number of \( i \) and \( i^{\prime} \) elements 
is assumed to be one each in this explanation, 
what is multiplied by 
( \textcircled{\scriptsize 3} - \textcircled{\scriptsize 2} )
is simply \( \partial V^{<1>} / \partial V^{<0>} \), which means 
how warped the mapping relationship is in {\sl Forward Subnet} 
on the basis of the Linear Case. 

Let us consider the case with \( \mbox{\textcircled{\scriptsize 3}} = 0.5 \) 
that corresponds to the center graph in the upper ones of Fig. 16.
In the Contractive Case with the green lines, 
a signal propagating through {\sl Backward Subnet} \textcircled{\scriptsize 4} 
becomes very small when \textcircled{\scriptsize 2} is around \( 0.5 \), 
since the value of ( \textcircled{\scriptsize 3} - \textcircled{\scriptsize 2} ) 
and inclination \( \partial V^{<1>} / \partial V^{<0>} \) are both small. 
In addition, it is important that 
the curve of \textcircled{\scriptsize 4} becomes an inflection point 
when \( \mbox{\textcircled{\scriptsize 2}} = 0.5 \). 
This produces a very unfavorable effect 
on the convergence dynamics of the whole network.
At the edges near \( 0.0 \) and \( 1.0 \) in \textcircled{\scriptsize 2}, 
on the other hand, the value of \textcircled{\scriptsize 4} 
sent to the internal space by {\sl Backward Subnet} is fairly large, 
since both ( \textcircled{\scriptsize 3} - \textcircled{\scriptsize 2} ) 
and \( \partial V^{<1>} / \partial V^{<0>} \) become large at the same time. 
In the Expansive Case with the orange lines, 
\( \partial V^{<1>} / \partial V^{<0>} \) is large around \( 0.5 \), 
but it is conversely small at the edges near \( 0.0 \) and \( 1.0 \).
The value of \textcircled{\scriptsize 4}, 
which is sent to the internal space by {\sl Backward Subnet}, 
is the product of \( \partial V^{<1>} / \partial V^{<0>} \) and 
( \textcircled{\scriptsize 3} - \textcircled{\scriptsize 2} ). 
As a result, \textcircled{\scriptsize 4} has two peaks 
between \( 0.0 \) and \( 0.5 \) and between \( 0.5 \) and \( 1.0 \). 
The value of \textcircled{\scriptsize 4} doesn't remain small 
even when \textcircled{\scriptsize 2} is close to \( 0.5 \), 
so {\sl Internetwork} in the Expansive Case 
seems to have an advantageous effect 
on the convergence dynamics of the whole network. 
These characteristics are basically common 
to the case of \( \mbox{\textcircled{\scriptsize 3}} \ne 0.5 \), 
and a signal transmitted to the internal space by {\sl Backward Subnet} 
varies in many ways, depending on 
whether {\sl Forward Subnet} has contractive or expansive mapping. 
In the Linear Case, \( \partial V^{<1>} / \partial V^{<0>} \) 
is a constant with value \( 1 \), 
so \textcircled{\scriptsize 4} is proportional to 
( \textcircled{\scriptsize 3} - \textcircled{\scriptsize 2} );
in the upper five graphs of Fig. 16, 
the purple lines are all straight. 

Figure 17 shows the time-course of the network states 
when the fixed value of \( 0.5 \) is applied 
as an input for the external space in an Association Mode, 
in which a pair of a feedback path and an input port 
exists only in the external space (Type \#1 in Fig. 3). 
In the same manner as in Fig. 15, 
the results are depicted as a table with three rows. 
The upper row indicates the Contractive Case 
with a mapping relationship that is obtained 
by the computation shown in Fig. 6, 
the middle one is the Linear Case 
obtained by the computation in Fig. 4, 
and the lower one shows the Expansive Case 
obtained by the computation in Fig. 7. 
For the horizontal direction, 
the arrangement of graphs is the same as that in Fig. 15. 
In this Association Mode, as stated above using Fig. 16, 
a signal flows through {\sl Backward Subnet}. 
Therefore, this behavior is fairly different from that of Fig. 15 
with a pair of a feedback path and an input port only in the internal space.

% Figure 17
\begin{figure}[!t]

    \hspace*{-6.5mm} 
    \includegraphics[scale=0.86]{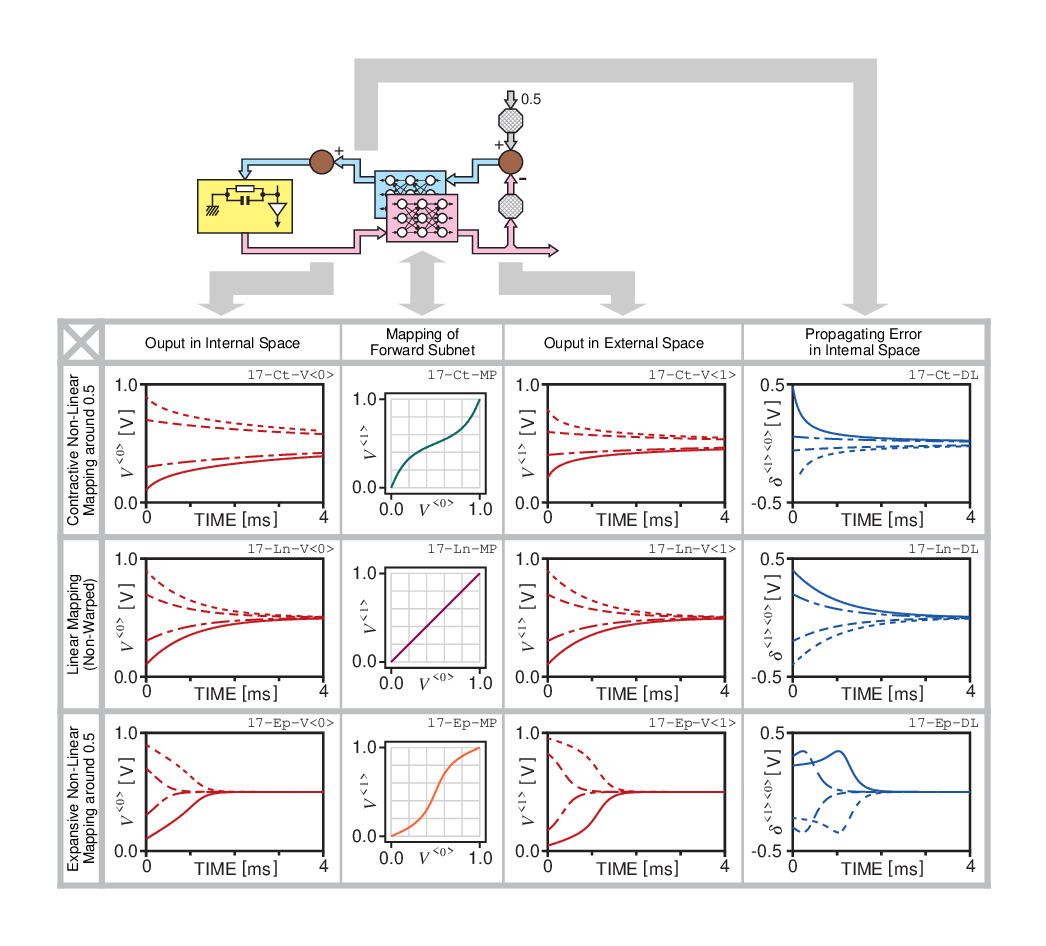}

    \vspace*{-3.0mm} 
    \caption{ 
      Simulation results for Type \#1 
      of the Association Mode in the one-dimensional model. 
      In this experiment, the place with a pair of a feedback path and an input port 
      is limited to the external space. 
      The graphs also include the Contractive, Linear, and Expansive Cases. 
    } 

\end{figure}

Notice the upper row of the Contractive Case. 
It is common to Fig. 15 that external {\tt 17-Ct-V<1>} 
is contractive overall compared with internal {\tt 17-Ct-V<0>}. 
When the output of the internal space in the early stage 
exists within the expansive mapping domain, 
the signal propagating through {\sl Backward Subnet} is rather large, 
but it becomes drastically smaller 
after the network state moves to the contractive mapping domain, 
and its convergence slows down as it approaches \( 0.5 \). 
Therefore, the effect that makes an error signal 
(an input from the outside - an output via a feedback connection) 
in the external space reduce in the internal space 
becomes smaller and smaller over time.
As a result, the output in the external space \( V^{<1>} \) 
converges very slowly toward \( 0.5 \).
We can also understand this phenomenon from the actual output 
of {\sl Backward Subnet} \( \delta^{<1><0>} \) as shown in {\tt 17-Ct-DL}. 

In the lower row of the Expansive Case, 
as in the lower row of Fig. 15, 
external {\tt 17-Ep-V<1>} is more expansive 
than internal {\tt 17-Ep-V<0>}. 
In this mapping case, produced is an effect 
that moves an error signal from the external space 
to the internal space and strongly reduces it there, 
since a signal flowing through {\sl Backward Subnet} 
retains its large value even around \( 0.5 \). 
As a result, both network states in the internal and external spaces 
converge quickly to the input value of \( 0.5 \). 
As indicated by the orange lines in Fig. 16, 
a signal propagating to the internal space via {\sl Backward Subnet} 
\textcircled{\scriptsize 4} is not monotonous but has two peaks 
between \( 0.0 \) and \( 0.5 \) and between \( 0.5 \) and \( 1.0 \). 
Interestingly, due to this, 
the decreasing behavior of the error signal \( \delta^{<1><0>} \) 
changes on the way, as shown in {\tt 17-Ep-DL}. 

In the middle row of the Linear Case, 
an input to and an output from {\sl Forward Subnet} are identical, 
and the dynamics in this Association Mode are also the same as
the case of Fig. 15 without any feedback signals 
via {\sl Forward Subnet} and {\sl Backward Subnet}.

% 4.2 Association Mode in Two-Dimensional Model 
\subsection{Association Mode in Two-Dimensional Model}

\noindent
In this subsection, we check the network behavior 
for the Association Mode in the two-dimensional model. 
Figure 18 shows the network's time-course 
with a pair of a feedback path and an input port 
only in the internal space (Type \#0 in Fig. 3), 
when fixed values \( (0.5, 0.5) \) are applied 
to the input port in the internal space. 
In the same way as in Figs. 15 and 17, 
the results are illustrated as a table with three rows. 
The upper row is for the network 
in the Contractive Case with {\sl Internetwork} 
obtained by the simulation shown in Fig. 11, 
the middle one is for the network 
in the Linear Case with that obtained in Fig. 9, and 
the lower one is for the network 
in the Expansive Case with that obtained in Fig. 13. 
For the horizontal direction, 
each of the left graphs is the time-course of the output in the internal space, 
its right one shows the mapping relationship 
already acquired in {\sl Internetwork}, 
and the rightmost one shows the time-course of the output in the external space. 
Here, we omit inputs and outputs to/from {\sl Backward Subnet} 
that were shown in the one-dimensional model. 
In the graphs, small black circles indicate the initial points, 
16 of which were selected for each case; 
all 16 initial points are arranged 
in such a manner that their positions in the internal spaces 
are common throughout the three cases. 
We chose five trajectories in each graph 
and drew ten small subsidiary arrows at most on each of those trajectories 
every \( 0.2 ~{\sf ms} \); 
a longer interval between the arrows means a faster change with time.

% Figure 18
\begin{figure}[!t]
  
    \hspace*{-4.0mm} 
    \includegraphics[scale=1.15]{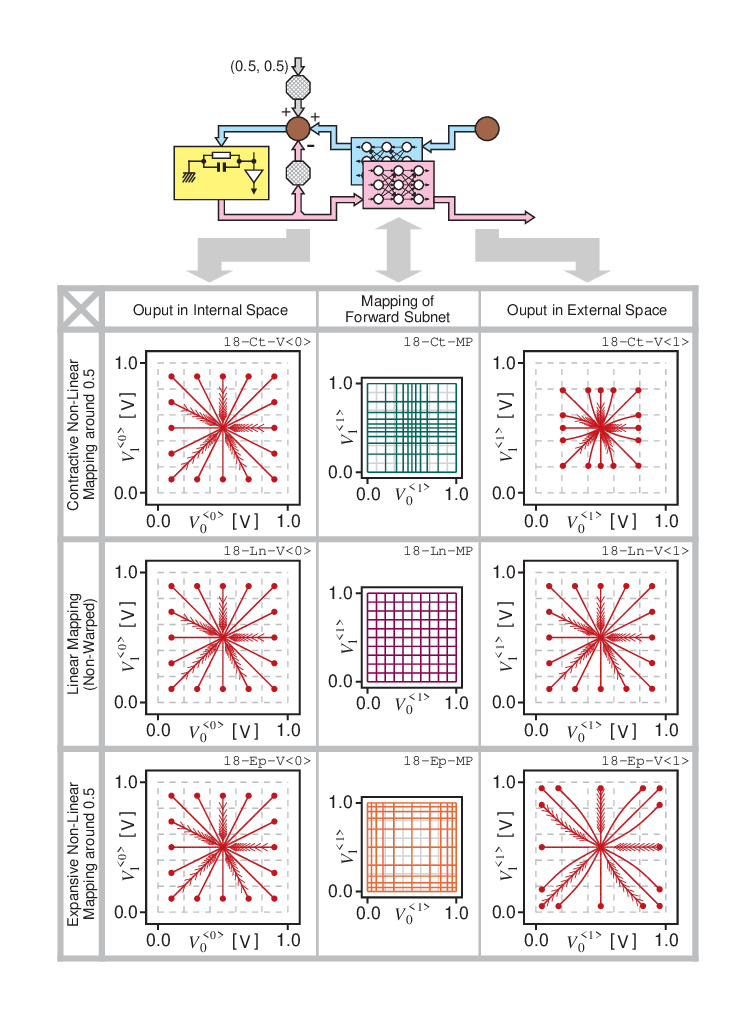}

    \vspace*{-5.0mm} 
    \caption{ 
      Simulation results for Type \#0 
      of the Association Mode in the two-dimensional model. 
      In this experiment, the place with a pair of a feedback path and an input port 
      is limited to the internal space. 
      The graphs include the Contractive, Linear, and Expansive Cases. 
    }

\end{figure}

For this Association Mode with a pair of a feedback path and an input port 
only in the internal space, 
as in the one-dimensional model, 
signals through {\sl Backward Subnet} do not exist. 
Therefore, the output in the internal space \( V^{<0>}_{i} \) 
is mapped simply and statically to that in the external space 
 \( V^{<1>}_{i'} \) by {\sl Forward Subnet}. 
Due to this, in the Linear Case shown in the middle row, 
the initial positions, the (straight) trajectories, and 
the convergent ways (intervals between arrows) of the outputs 
are identical in the internal and external spaces. 

How is the Contractive Case shown in the upper row? 
On the basis of the case shown in the middle row, 
we summarize its convergence characteristics as follows:

\begin{itemize}
  \setlength{\leftskip}{4mm}
  \setlength{\labelsep}{2mm}

  \item[(0-a)] The initial points in the external space are located 
               nearer to the center \( (0.5, 0.5) \) 
               than those in the internal space. 

  \item[(0-b)] When the initial points are on a diagonal line 
               connecting \( (0.0, 0.0) \) and \( (1.0, 1.0) \), 
               or on that between \( (0.0, 1.0) \) and \( (1.0, 0.0) \), 
               \( V^{<1>}_{0} \) and \( V^{<1>}_{1} \) vary 
               in the same manner, 
               and so the network state converges 
               straightforwardly to the center. 

  \item[(0-c)] When the initial points are on a straight line 
               between \( (0.0, 0.5) \) and \( (1.0, 0.5) \), 
               or on that connecting \( (0.5, 0.0) \) and \( (0.5, 1.0) \), 
               the varing tendencies of \( V^{<1>}_{0} \) and \( V^{<1>}_{1} \) 
               are different from each other, 
               but the network state converges straightforwardly to the center, 
               since one parameter already remains in a converging state 
               and only the other parameter varies. 

  \item[(0-d)] As for the other initial positions 
               except for the case of (0-b) and (0-c), 
               \( V^{<1>}_{0} \) and \( V^{<1>}_{1} \) vary in different ways, 
               and the network state converges in a curved path to the center.
               Then the curved line bends in the direction with a steep gradient, 
               in such a manner that the network state 
               gets closer to a neighbor straight line \( V^{<1>}_{1} = V^{<1>}_{0} \) 
               or \( V^{<1>}_{1} = - V^{<1>}_{0} + 1.0 \). 

  \item[(0-e)] Overall, the output in the external space {\tt 18-Ct-V<1>} 
               is a reduced-size version of {\tt 18-Ct-V<0>} 
               according to the mapping relationship of {\sl Forward Subnet}. 
\end{itemize}

In the Expansive Case shown in the lower row, 
(0-b) and (0-c) are shared, but the behavior in (0-a), (0-d), and (0-e) 
is different from that in this case as follows:

\begin{itemize}
  \setlength{\leftskip}{5mm}
  \setlength{\labelsep}{2mm}

  \item[(0-a')] The initial positions in the external space 
                are located farther from the center 
                compared with the Contractive and Linear Cases. 

  \item[(0-d')] As for the other initial positions 
                except for the cases in (0-b) and (0-c), 
                \( V^{<1>}_{0} \) and \( V^{<1>}_{1} \) vary in different ways, 
                and the network state converges in a curved path to the center. 
                But their curving tendencies are opposite to those 
                in the Contractive Case of the upper row; 
                the network state approaches 
                a straight line \( V^{<1>}_{0} = 0.5 \) or \( V^{<1>}_{1} = 0.5 \). 

  \item[(0-e')] Contrary to the Contractive Case, 
                the output in the external space {\tt 18-Ep-V<1>} 
                is an extended-size version of {\tt 18-Ep-V<0>} 
                based on the mapping relationship of {\sl Forward Subnet}.
                The intervals between arrows in {\tt 18-Ep-V<1>} 
                are almost equal before \( 2 ~{\sf ms} \), 
                because the convergence speed in the edges 
                near \( 0.0 \) and \( 1.0 \) 
                is suppressed due to the contractive mapping relationship in those domains. 
                This phenomenon can easily be imagined 
                from the straightforward variation 
                during \( 0 ~{\sf ms} - 2 ~{\sf ms} \)
                in the one-dimensional model of {\tt 15-Ep-V<1>}. 
\end{itemize}

On the other hand, Fig. 19 illustrates the behavior of a network 
with a pair of a feedback path and an input port 
only in the external space (Type \#1 in Fig. 3), 
when fixed values \( (0.5, 0.5) \) are applied 
to the input port in the external space. 
The results are summarized as three rows as in Fig. 18. 
The upper row means the Contractive Case with {\sl Internetwork} 
obtained through the computation in Fig. 11, 
the middle one is the Linear Case with that in Fig. 9, 
and the lower one is the Expansive Case with that in Fig. 13. 
For the horizontal direction, 
the arrangement of graphs is the same as that in Fig. 18. 
The 16 initial positions in each case are set 
to be equivalent in the internal space. 
The information for the inputs and outputs to/from {\sl Backward Subnet} 
is also omitted in the figure. 
Since the signals through {\sl Backward Subnet} exist in this Association Mode, 
the state variation in the external space 
should affect that in the internal space. 
First, let us check the Linear Case in the middle row. 
The information for the feedback and the input in the external space 
is sent to the internal space as it is. 
So in the one-dimensional model, 
{\tt 15-Ln-V<0>} and {\tt 15-Ln-V<1>} in the middle row of Fig. 15 were the same, 
as were {\tt 17-Ln-V<0>} and {\tt 17-Ln-V<1>} in the middle row of Fig. 17. 
In the two-dimensional model, 
{\tt 19-Ln-V<0>} and {\tt 19-Ln-V<1>} are identical to each other 
and are equal to {\tt 18-Ln-V<0>} and {\tt 18-Ln-V<1>} of Fig. 18.

% Figure 19
\begin{figure}[!t]

    \hspace*{-4.0mm} 
    \includegraphics[scale=1.15]{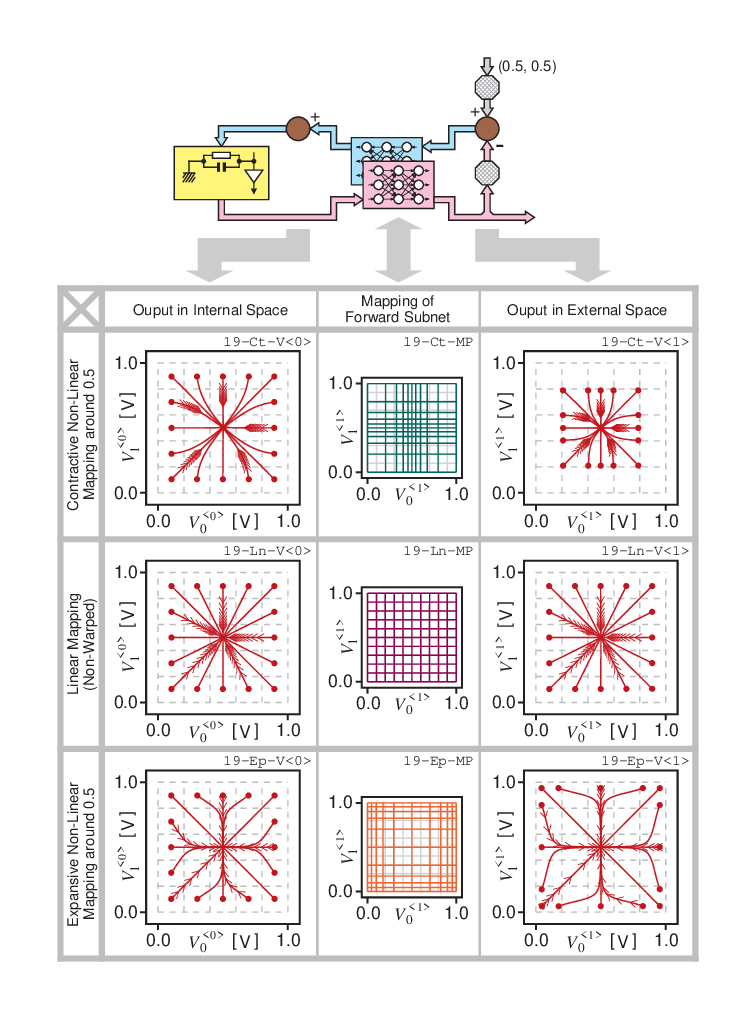}

    \vspace*{-5.0mm} 
    \caption{ 
      Simulation results for Type \#1 
      of the Association Mode in the two-dimensional model. 
      In this experiment, the place with a pair of a feedback path and an input port 
      is limited to the external space. 
      The graphs also include the Contractive, Linear, and Expansive Cases. 
    }

\end{figure}

Next, on the basis of these results in the middle rows, 
we check the Contractive Case in the upper row. 
The convergence characteristics can be itemized as follows:

\begin{itemize}
  \setlength{\leftskip}{4mm}
  \setlength{\labelsep}{2mm}

  \item[(1-a)] The initial positions in the external space are located 
               nearer to the center as with (0-a). 

  \item[(1-b)] When an initial position is located on a diagonal line 
               connecting \( (0.0, 0.0) \) with \( (1.0, 1.0) \), 
               or on that between \( (0.0, 1.0) \) and \( (1.0, 0.0) \), 
               the network state converges linearly to the center
               since \( V^{<1>}_{0} \) and \( V^{<1>}_{1} \) vary 
               with the same tendency.

  \item[(1-c)] When an initial point is located 
               on a straight line between \( (0.0, 0.5) \) and \( (1.0, 0.5) \), 
               or on that between \( (0.5, 0.0) \) and \( (0.5, 1.0) \), 
               the network state also converges linearly to the center 
               since either \( V^{<1>}_{0} \) or \( V^{<1>}_{1} \) 
               has already converged on a fixed value and the other only varies. 

  \item[(1-d)] Except for the initial points in (1-b) and (1-c), 
               the network state goes to the center in a curved line, 
               since \( V^{<1>}_{0} \) and \( V^{<1>}_{1} \) vary 
               with mutually different tendencies  
               based on the mapping relationship of {\sl Forward Subnet}; 
               the network state varies toward the center 
               in such a manner that it approaches a diagonal line where 
               \( V^{<1>}_{1} = V^{<1>}_{0} \) or 
               \( V^{<1>}_{1} = - V^{<1>}_{0} + 1.0 \). 
               Then the variation occurring in the external space is 
               returned to the internal space by {\sl Backward Subnet}, 
               and further transmitted to the external space 
               through {\sl Forward Subnet}. 
               Due to this looping, the convergence trajectory 
               in the internal space curves, 
               and that in the external space curves more; 
               this result is obvious 
               by comparing {\tt 19-Ct-V<0>} and {\tt 19-Ct-V<1>} 
               with {\tt 18-Ct-V<0>} and {\tt 18-Ct-V<1>} of Fig. 18. 

  \item[(1-e)] It is common to Fig. 18 that the network state in the external space 
               is contractive overall compared with that in the internal space.
               But as the output in the external space approaches the center, 
               the corresponding error signal through {\sl Backward Subnet} 
               becomes smaller in this case. 
               So an effect, which suppresses the error information 
               of the external space inside the internal space, 
               is not strongly produced; 
               this phenomenon is basically common to the relation between 
               {\tt 17-Ct-V<0>} and {\tt 17-Ct-V<1>} in the one-dimensional model. 
               Therefore, the speed with which 
               the outputs \( V^{<1>}_{0} \) and \( V^{<1>}_{1} \) approach the center 
               is very 
% 
%%%%% ----- Updated Part No. 02 (Begin) ----- 
%%%%% v1 
%                is very slow, 
%%%%% v2 
               low, 
%%%%% ----- Updated Part No. 02 (End) ----- 
% 
               and the time necessary for reaching the center becomes very large. 
               Thus, in {\tt 19-Ct-V<0>} and {\tt 19-Ct-V<1>}, 
               the intervals between small subsidiary arrows on the trajectories 
               become much narrower in such a stage that remains far from the center. 
\end{itemize}

In the Expansive Case illustrated in the lower row, 
the network state's variation seems very different 
from that in the Contractive Case. 
There also appears to be a big difference in the dynamical properties 
between Types \#0 and \#1. 
The convergence characteristics stated in (1-b) and (1-c) 
are the same as those in the Expansive Case, 
but the behavior in (1-a), (1-d) and (1-e) differs 
from that in this case:

\begin{itemize}
  \setlength{\leftskip}{5mm}
  \setlength{\labelsep}{2mm}

  \item[(1-a')] The initial points in the external space are located 
                farther from the center as with (0-a').

  \item[(1-d')] Except for the initial positions in (1-b) and (1-c), 
                the varying tendencies of \( V^{<1>}_{0} \) and \( V^{<1>}_{1} \) 
                are different from each other, 
                so the network state also approaches 
                the center in a curved line in this case. 
                But note that the curving direction is opposite 
                to that in the Contractive Case in the upper row; 
                the network state follows a curve 
                that becomes a straight line where 
                \( V^{<1>}_{0} = 0.5 \) or \( V^{<1>}_{1} = 0.5 \). 
                Based on the looping 
                by {\sl Backward Subnet} and {\sl Forward Subnet}, 
                the trajectory in the internal space curves and 
                that in the external space curves more; 
                this result is clarified by comparing 
                {\tt 19-Ep-V<0>} and {\tt 19-Ep-V<1>} with 
                {\tt 18-Ep-V<0>} and {\tt 18-Ep-V<1>} of Fig. 18. 

  \item[(1-e')] For this mapping case, as stated using Fig. 16, 
                a signal through {\sl Backward Subnet} 
                stays large even near the center, 
                so an effect, which suppresses the error information 
                for the external space 
                inside the internal space, is strongly produced. 
                Due to this, the network state goes to the center very quickly. 
                The intervals between the small subsidiary arrows 
                in {\tt 19-Ep-V<0>} and {\tt 19-Ep-V<1>}, 
                which are wider overall than those 
                in {\tt 18-Ep-V<0>} and {\tt 18-Ep-V<1>} of Fig. 18, 
                also suggest fast convergence. 
\end{itemize}

% 5 Discussion 
\section{Discussion}

% 5.1 Constrained Association Mode 
\subsection{Constrained Association Mode}

\noindent
In the previous section, 
we investigated the dynamical behavior of networks 
defined as the Association Mode in which 
a pair of a feedback path and an input port 
is restricted in either the internal or external space. 
In an Association Mode in which a pair of a feedback path and an input port 
exists only in the external space (Type \#1 in Fig. 3), 
we confirmed interesting phenomena 
where the convergence trajectories 
emerged from straight lines in both spaces, 
depending on the differences 
between the inputs and the present outputs in the external space, 
\( (Z^{<1>}_{0} - V^{<1>}_{0}) \) and \( (Z^{<1>}_{1} - V^{<1>}_{1}) \), 
and the warping in the mapping relationship of {\sl Internetwork}. 

If we put a trajectory toward a goal point to the input port 
instead of only applying the goal point there, 
what is the network behavior of each architecture in the Association Mode?
It is also interesting to examine how the network dynamics vary 
depending on the speed of the target trajectory 
that is applied to the input port. 
In the Type \#1 architecture of the Association Mode, for instance, 
as the speed of a target trajectory applied to the input port increases, 
the difference between the input and the present output 
in the external space will increase, 
and the input and the output to/from {\sl Backward Subnet}, 
 \( \delta^{<1><3>}_{i} \) and \( \delta^{<1><0>}_{i} \), will become larger. 
Since this situation is basically similar to the Association Mode 
discussed in the previous section, 
we simply expect that the convergence trajectories curve largely 
depending on the warping in the mapping relationship of {\sl Internetwork}. 
In contrast, as the speed of the target trajectory decreases, 
\( \delta^{<1><3>}_{i} \) and \( \delta^{<1><0>}_{i} \) will become smaller 
due to the function of dynamical neurons in the internal space. 
So the convergence trajectory's curve in the external space may become smaller, 
even if warping in the mapping relationship of {\sl Internetwork} is great. 
We name this particular Association Mode, in which a target trajectory 
instead of a fixed goal point is put to the input port, 
the ``Constrained" Association Mode, 
and investigate its dynamical behavior. 
In the following simulation studies, we assume 
three speeds (time constants), as shown in Fig. 20, 
where the moving times required from an initial point 
to a goal point (the center) are denoted. 
Eq. (29) is set to OFF in the same way as in Section 4. 
All the conditions except for these ones 
are common to those in Sections 3 and 4.

% Figure 20
\begin{figure}[!t]

    \hspace*{-6.5mm} 
% 
%%%%% ----- Updated Part No. 03 (Begin) ----- 
%%%%% v1 
%     \includegraphics[scale=0.85]{./tsutsumi_niebur_01_fig_20.eps}
%%%%% v2 
    \includegraphics[scale=0.85]{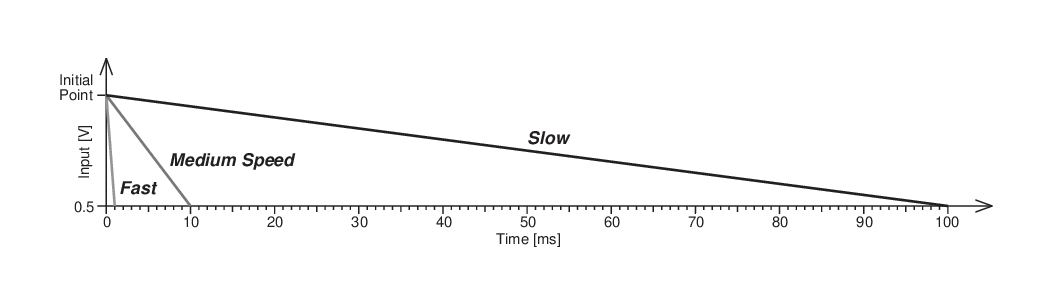}
%%%%% ----- Updated Part No. 03 (End) ----- 
% 

    \vspace*{-3.0mm} 
    \caption{ 
      Three target trajectories 
      with different speeds (time constants) 
      employed in two types of the Constrained Association Modes. 
    } 

\end{figure}

% 5.2 Type \#0 of Constrained Association Mode 
\subsection{Type \#0 of Constrained Association Mode}

\noindent
Figure 21 shows the time-course of the network dynamics 
with a pair of a feedback path and an input port 
only in the internal space (Type \#0 in Fig. 3) 
when a straight target trajectory is applied to the input port.
The whole is divided into upper and lower blocks, 
which respectively show the Contractive and Expansive Cases. 
Each block has three rows, which are graphs 
depicted based on the simulation results 
with 
% 
%%%%% ----- Updated Part No. 04 (Begin) ----- 
%%%%% v1 
% fast, medium, and slow speeds 
%%%%% v2 
high, medium, and low speeds 
%%%%% ----- Updated Part No. 04 (End) ----- 
% 
of the target trajectory from the top. 
The leftmost graphs show the target trajectories 
applied to the input port in the internal space.
Their right graphs are the outputs in the internal space, 
and the next right ones depict the mapping relationship of {\sl Forward Subnet} 
to increase intuitive understanding. 
The rightmost graphs are the outputs in the external space. 
In this figure, we selected five trajectories and 
depicted ten small subsidiary arrows at most on each of those trajectories
at every particular period of time from the initial state 
to visually grasp their convergence speeds; 
the intervals among the small subsidiary arrows are 
\( 0.2 {\sf ms} \), \( 2 {\sf ms} \), and \( 20 {\sf ms} \) 
from the top in each block. 
We omitted the Linear Case, since, as seen in the previous section, 
the outputs in the internal and external spaces are exactly the same 
and the trajectories are straight 
in this architecture of the Association Mode.

% Figure 21
\begin{figure}[p]

    \vspace*{-3.0mm}
    \hspace*{6.5mm}
% 
%%%%% ----- Updated Part No. 10 (Begin) ----- 
%%%%% v1 
%     \includegraphics[scale=0.77]{./tsutsumi_niebur_01_fig_21.eps}
%%%%% v2 
    \includegraphics[scale=0.77]{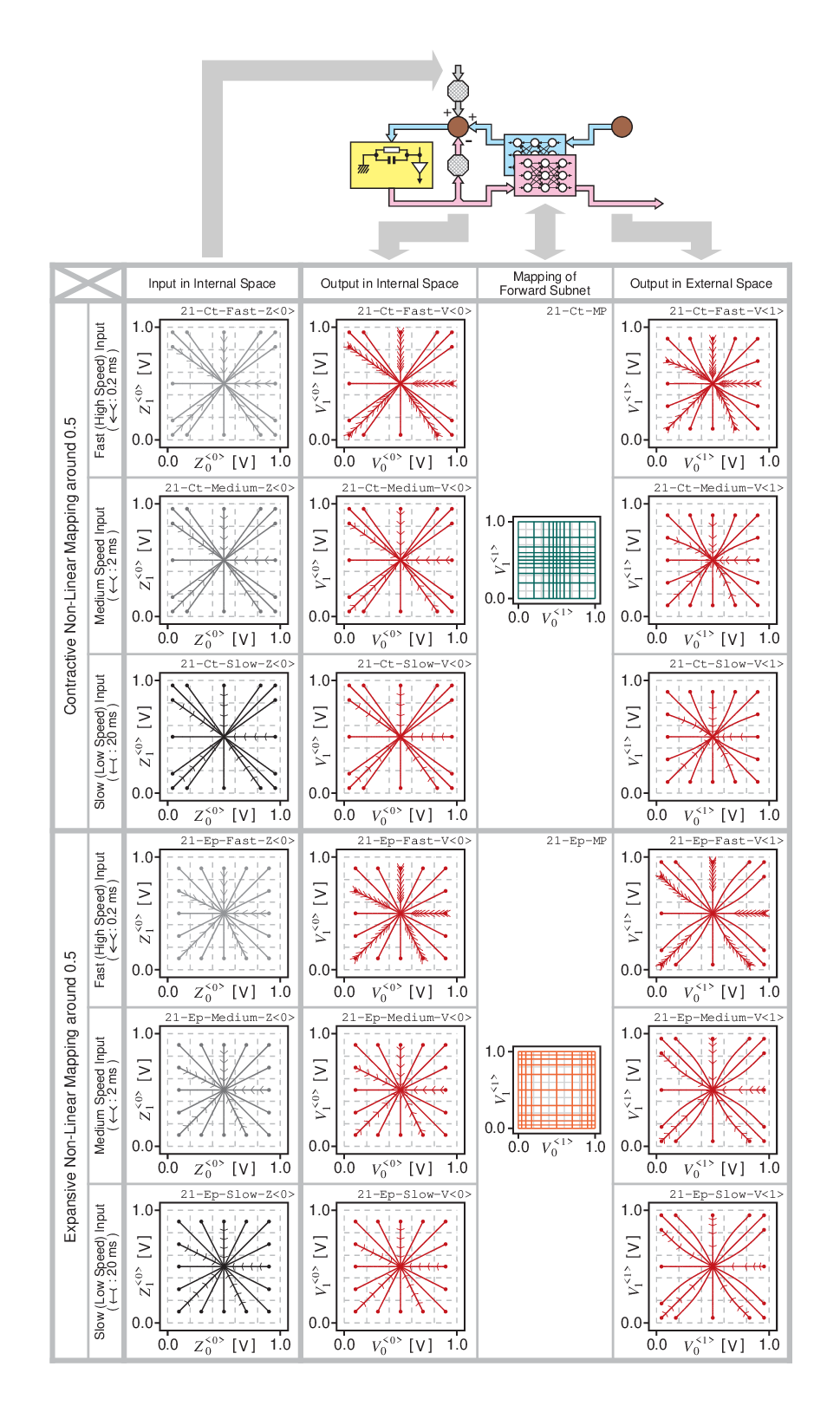}
%%%%% ----- Updated Part No. 10 (End) ----- 
% 

    \vspace*{-5.0mm}
    \caption{ 
      Simulation results for Type \#0 
      of the Constrained Associative Mode in the two-dimensional model. 
      In this experiment, 
      a straight target trajectory instead of a goal point 
      is applied to the input port existing only in the internal space. 
      Network behavior for three different speeds of target trajectories 
      is illustrated. 
      The graphs only include the Contractive and Expansive Cases. 
    } 

\end{figure}

First, we consider the upper block for the Contractive Case. 
For a comparison with the experimental results stated later, 
the initial positions in the internal space are set 
to those farther from the center, 
in such a manner that the initial positions in the external space 
equal those in the internal space shown in Section 4. 
Since there is no signal through {\sl Backward Subnet} in this case, 
the internal space's output straightforwardly approaches the goal point, 
and an arbitrary point on the straight output trajectory in the internal space 
is mapped simply in a static way to that in the external space. 
 {\tt 21-Ct-Fast-V<0>} and {\tt 21-Ct-Fast-V<1>} with 
% 
%%%%% ----- Updated Part No. 05 (Begin) ----- 
%%%%% v1 
% fast speed 
%%%%% v2 
fast (high speed) 
%%%%% ----- Updated Part No. 05 (End) ----- 
% 
inputs 
are basically the same as {\tt 18-Ct-V<0>} and {\tt 18-Ct-V<1>} 
in the previous section, 
although the similarity is somewhat obscure at a glance 
since the initial positions in both cases are different. 
Even if the speed of a target trajectory decreases, 
the fact remains that an arbitrary point on the straight output trajectory 
in the internal space is mapped statically to that in the external space. 
Comparisons of {\tt 21-Ct-Fast-V<1>}, {\tt 21-Ct-Medium-V<1>}, 
and {\tt 21-Ct-Slow-V<1>} confirm that 
the output in the external space draws the same trajectory 
regardless of whether the target trajectory's speed is 
% 
%%%%% ----- Updated Part No. 06 (Begin) ----- 
%%%%% v1 
% fast or slow. 
%%%%% v2 
high or low. 
%%%%% ----- Updated Part No. 06 (End) ----- 
% 

The Expansive Case is shown in the lower block. 
The initial positions in the internal space are set 
to the same ones as in Figs. 18 and 19 of Section 4. 
Also in this case, 
an arbitrary point on a straight output trajectory in the internal space 
is simply mapped in a static manner to that in the external space. 
{\tt 21-Ep-Fast-V<0>} and {\tt 21-Ep-Fast-V<1>} with 
% 
%%%%% ----- Updated Part No. 07 (Begin) ----- 
%%%%% v1 
% fast speed 
%%%%% v2 
fast (high speed) 
%%%%% ----- Updated Part No. 07 (End) ----- 
% 
inputs 
are the same as {\tt 18-Ep-V<0>} and {\tt 18-Ep-V<1>} in the Association Mode.
As is clear from the comparison of
{\tt 21-Ep-Fast-V<1>}, {\tt 21-Ep-Medium-V<1>}, and {\tt 21-Ep-Slow-V<1>}, 
the output trajectories in the external space are almost the same 
regardless of whether the target trajectory's speed is 
% 
%%%%% ----- Updated Part No. 08 (Begin) ----- 
%%%%% v1 
% fast or slow. 
%%%%% v2 
high or low. 
%%%%% ----- Updated Part No. 08 (End) ----- 
% 
Note that the output curving in the upper block 
is opposite to that in the lower block; 
this fact is common to the results for Type \#0 of the Association Mode.

% 5.3 Type \#1 of Constrained Association Mode 
\subsection{Type \#1 of Constrained Association Mode}

\noindent
Figure 22 shows the time-course of the network dynamics 
with a pair of a feedback path and an input port 
only in the external space (Type \#1 in Fig. 3) 
when a straight target trajectory is applied to the input port. 
As in Fig. 21, three rows in the upper block mean the Contractive Case, 
and those in the lower block indicate the Expansive Case. 
The Linear Case was omitted 
for the same reason as in the previous subsection. 
The three rows in each block respectively denote the results with 
% 
%%%%% ----- Updated Part No. 09 (Begin) ----- 
%%%%% v1 
% fast, medium, and slow speeds 
%%%%% v2 
high, medium, and low speeds 
%%%%% ----- Updated Part No. 09 (End) ----- 
% 
of the target trajectory from the top; 
these three speeds of the target trajectory are common to those in Fig. 21. 
The rightmost graphs are the target trajectories 
applied to the input port in the external space. 
The outputs in the internal and external spaces based on these inputs 
are shown in order from the left, 
placing the mapping relationship in {\sl Forward Subnet} between them.

% Figure 22
\begin{figure}[p]

    \vspace*{-3.0mm}
    \hspace*{6.5mm}
% 
%%%%% ----- Updated Part No. 11 (Begin) ----- 
%%%%% v1 
%     \includegraphics[scale=0.77]{./tsutsumi_niebur_01_fig_22.eps}
%%%%% v2 
    \includegraphics[scale=0.77]{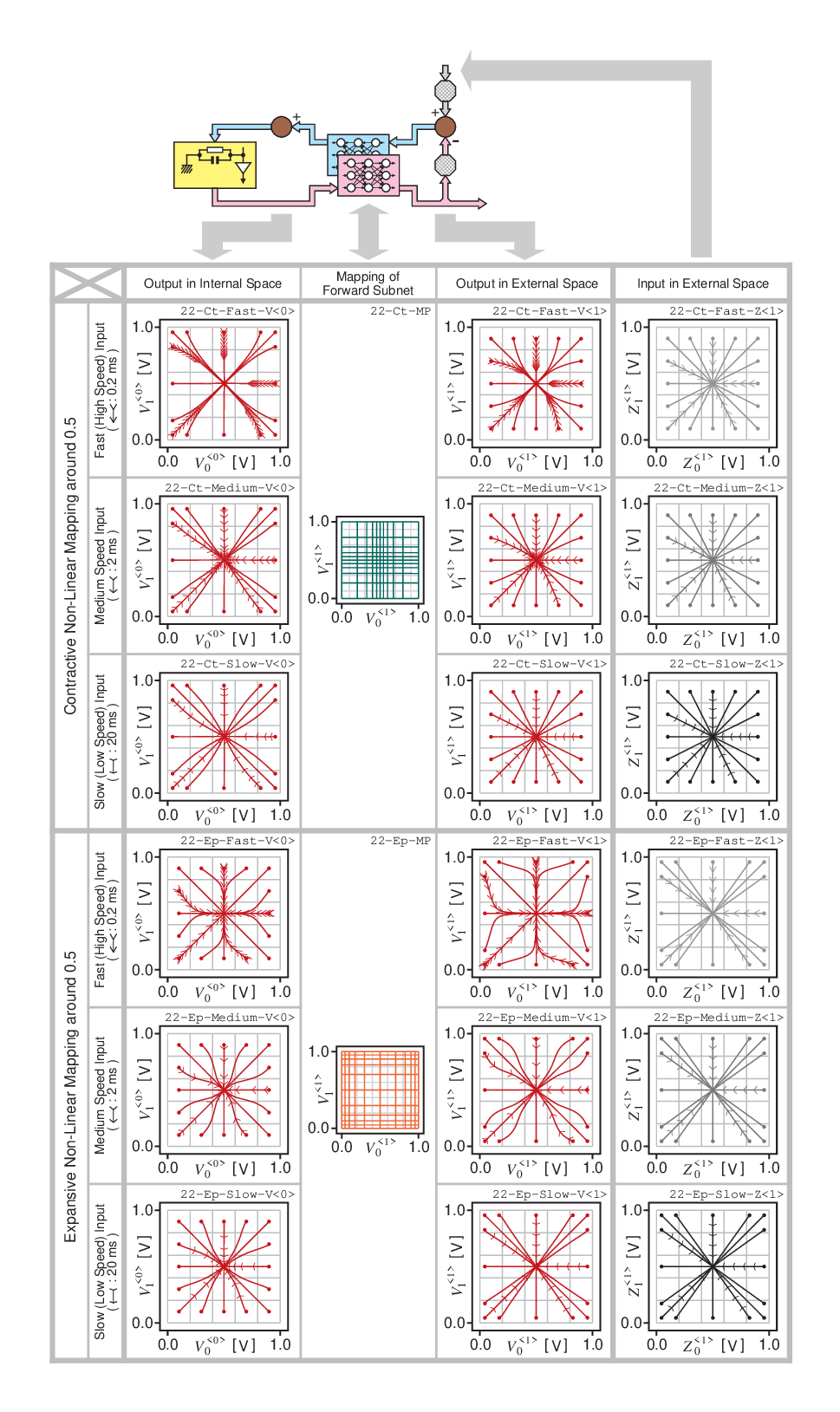}
%%%%% ----- Updated Part No. 11 (End) ----- 
% 

    \vspace*{-5.0mm}
    \caption{ 
      Simulation results for Type \#1 
      of the Constrained Associative Mode in the two-dimensional model. 
      In this experiment, 
      a straight target trajectory instead of a goal point 
      is applied to the input port existing only in the external space. 
      Network behavior for three different speeds of target trajectories 
      is depicted. 
      The graphs also only include the Contractive and Expansive Cases. 
    } 

\end{figure}

First, we look at the Contractive Case in the upper block. 
The uppermost graphs {\tt 22-Ct-Fast-V<0>} and {\tt 22-Ct-Fast-V<1>} 
show basically the same responses as 
the uppermost graphs of Fig. 19 in the Association Mode 
{\tt 19-Ct-V<0>} and {\tt 19-Ct-V<1>}, 
although their initial positions are different. 
In both situations, large signals flow through {\sl Backward Subnet} 
depending on the mapping relationship in {\sl Internetwork}. 
Due to a detoured feedback loop via the external space, 
they affect both the internal and external spaces, 
where the output trajectories curve more greatly away from the straight. 
Those graphs with a fast target trajectory 
have characteristics common with the results in the Association Mode, 
but intriguingly, the graphs in the middle and lower rows 
show different tendencies from those in the upper row.
By comparing {\tt 22-Ct-Medium-V<0>} and {\tt 22-Ct-Medium-V<1>} 
in the middle row with {\tt 22-Ct-Fast-V<0>} and {\tt 22-Ct-Fast-V<1>} in the upper row, 
the former curving for the trajectories in both the internal and external spaces 
is overall milder than the latter one. 
For the case with an even slower target trajectory shown in the lower row, 
the output in the external space {\tt 22-Ct-Slow-V<1>} draws a straight line 
and nearly corresponds to the target trajectory {\tt 22-Ct-Slow-Z<1>}. 
Instead, the output in the internal space \( ( V^{<0>}_{0}, V^{<0>}_{1} ) \) 
draws a curved trajectory. Note that 
its curving direction is opposite to that with a fast target trajectory; 
it is obvious in comparison of {\tt 22-Ct-Slow-V<0>} with {\tt 22-Ct-Fast-V<0>}. 

The relationship between Figs. 21 and 22 is also crucial. 
Compare the graphs in the lower row in the lower block of Fig. 21 
with those in the lower row in the upper block of Fig. 22.
Significantly, the output trajectories 
in the internal and external spaces are opposite to each other 
in these two cases with the slow target trajectories in Figs. 21 and 22 
in such a manner that {\tt 21-Ep-Slow-V<0>} and {\tt 22-Ct-Slow-V<1>} are the same, 
as are {\tt 21-Ep-Slow-V<1>} and {\tt 22-Ct-Slow-V<0>}. 
Since the network dynamics are observed for a sufficiently long time 
and their results are plotted in Figs. 21 and 22, 
all of the output trajectories go to the goal point (the center). 
As predicted from the results in the one-dimensional model of the Assoiciation Mode, 
{\tt 17-Ct-V<0>} and {\tt 17-Ct-V<1>} of Fig. 17, 
the convergence speed near the center is very 
% 
%%%%% ----- Updated Part No. 12 (Begin) ----- 
%%%%% v1 
% slow 
%%%%% v2 
low 
%%%%% ----- Updated Part No. 12 (End) ----- 
% 
in the Contractive Case. 
For this reason, the intervals between small subsidiary arrows 
on a trajectory become dense
in {\tt 22-Ct-Fast-V<0>} or {\tt 22-Ct-Fast-V<1>}, for instance. 
This phenomenon is caused regardless of the speed of a target trajectory 
applied to the input port, and so the convergence speed near the center 
is very 
% 
%%%%% ----- Updated Part No. 13 (Begin) ----- 
%%%%% v1 
% slow 
%%%%% v2 
low 
%%%%% ----- Updated Part No. 13 (End) ----- 
% 
even in {\tt 22-Ct-Slow-V<0>} and {\tt 22-Ct-Slow-V<1>}. 
Therefore, even though the convergence trajectories are identical, 
the periods required for converging to a goal point are not necessarily the same;
this point is also a critical phenomenon 
that depends on the mapping relationship in {\sl Internetwork}. 

Next, we investigate the Expansive Case in the lower block. 
When a fast target trajectory is applied to the input port, 
as stated with respect to Fig. 19 in Type \#1 of the Association Mode, 
large signals go through {\sl Backward Subnet} 
depending on the mapping relationship in {\sl Internetwork}. 
They affect both the internal and external spaces 
due to a detoured feedback loop via the external space, and 
the output trajectories curve greatly.
By comparing {\tt 22-Ep-Fast-V<0>} and {\tt 22-Ep-Fast-V<1>} 
with {\tt 19-Ep-V<0>} and {\tt 19-Ep-V<1>} of Fig. 19 in the Association Mode, 
these pairs are almost the same. 
Those graphs with a fast target trajectory 
have characteristics common to the results in the Association Mode, 
but it is interesting that the graphs in the middle and lower rows 
indicate tendencies different from those in the upper row.
In comparison of {\tt 22-Ep-Medium-V<0>} and {\tt 22-Ep-Medium-V<1>} 
in the middle row with {\tt 22-Ep-Fast-V<0>} and {\tt 22-Ep-Fast-V<1>} in the upper one, 
the former curving for the trajectories in both the internal and external spaces 
becomes milder than the latter one. 
Since this mapping relationship is expansive around the center 
but contractive at the edges near \( 0.0 \) and \( 1.0 \), 
we confirmed that the curving in a output trajectory is not monotonous 
but its tendency varies on the way to the center. 
For the case with an even slower target trajectory shown in the lower row, 
the output in the external space {\tt 22-Ep-Slow-V<1>} draws a straight line 
and nearly corresponds to the target trajectory {\tt 22-Ep-Slow-Z<1>} 
due to the function of the dynamical neurons in the internal space. 
Instead, the output in the internal space \( V^{<0>}_{i} \) draws a curved trajectory, 
but its curving direction is opposite to that with a fast target trajectory; 
this observation is clear in the comparison of 
{\tt 22-Ep-Slow-V<0>} with {\tt 22-Ep-Fast-V<0>}. 

Also in this case, we noticed an interesting relationship between Figs. 21 and 22.
In a comparison of the graphs in the lower row in the upper block of Fig. 21 
with those in the lower row in the lower block of Fig. 22, 
the output trajectories in the internal and external spaces are opposite to each other 
in the two cases with slow target trajectories of Figs. 21 and 22 
in such a manner that {\tt 21-Ct-Slow-V<0>} and {\tt 22-Ep-Slow-V<1>} are the same, 
as are {\tt 21-Ct-Slow-V<1>} and {\tt 22-Ep-Slow-V<0>}.

% 5.4 On Inverse Mapping 
\subsection{On Inverse Mapping}

\noindent
For Type \#0 of the Constrained Association Mode, 
the relationship between the output trajectories 
in the internal and external spaces is common 
regardless of whether a fast or a slow target trajectory 
is applied to the input port. 
A straight target trajectory in the internal space was transformed 
based on the mapping relationship already acquired in {\sl Internetwork}, 
and its warped trajectory simply appeared in the external space 
as a statically mapped path. 
In contrast, for Type \#1 of the Constrained Association Mode, 
a different appearance was presented. 
When a fast target trajectory was applied to the input port, 
the output trajectories in both the internal and external spaces 
are determined to emphasize their curving tendency 
due to a detoured feedback loop via the external space. 

When, in Type \#1 architecture, 
a very slow target trajectory from an initial point to a goal point 
is applied to the input port in the external space, 
conversely, the output in the external space 
completely follows its target trajectory; 
no signal flows through {\sl Backward Subnet}. 
If the mapping relationship from the internal space 
to the external space by {\sl Forward Subnet} is now assumed to be \( M \), 
an output trajectory, which was converted from that in the external space 
by the ``inverse" mapping relationship \( M^{-1} \), 
appears in the internal space in such a manner that 
the output in the external space transmitted by {\sl Forward Subnet} 
simply becomes a straight target trajectory 
that is applied to the input port in the external space. 
The following are specific examples:
The mapping from the output in the internal space {\tt 21-Ct-Slow-V<0>} 
to that in the external space {\tt 21-Ct-Slow-V<1>} 
in the Contractive Case for Type \#0 of the Constrained Association Mode 
is the same as 
the mapping from that in the external space {\tt 22-Ep-Slow-V<1>} 
to that in the internal space {\tt 22-Ep-Slow-V<0>} 
in the Expansive Case for Type \#1 of the Constrained Association Mode. 
In addition, 
the mapping from the output in the internal space {\tt 21-Ep-Slow-V<0>} 
to that in the external space {\tt 21-Ep-Slow-V<1>} 
in the Expansive Case for Type \#0 of the Constrained Association Mode 
is the same as 
the mapping from that in the external space {\tt 22-Ct-Slow-V<1>} 
to that in the internal space {\tt 22-Ct-Slow-V<0>} 
in the Contractive Case for Type \#1 of the Constrained Association Mode. 

Thus, when an extremely slow target trajectory 
is applied to the input port in the external space 
on Type \#1 of the Constrained Association Mode, 
an output trajectory, 
which was converted from the output in the external space 
by the inverse mapping \( M^{-1} \), 
appears in the internal space, 
although only the mapper from the internal space 
to the external space \( M \) does exist 
but the one for the inverse mapping \( M^{-1} \) 
does not exist inside the whole network.

% 6 Conclusions 
\section{Conclusions}

\noindent
In this paper, we propose a hierarchically modular neural network model 
composed of two types of neurons with different time constants 
(one is dynamic and the other is regarded as static), 
and precisely examined its dynamical behavior through simulation studies. 
We summarize our results in the following.

\vspace{4mm} 
\noindent 
\( \diamondsuit \) Model Proposal

\begin{itemize}

  \setlength{\itemsep}{-0.05mm}

  \item Supposing that the outputs of dynamical neurons are transformed 
        to other output variables by a mapping function or 
        a multi-layered neural network with static neurons, 
        we designed an energy function consisting of those output variables. 
        A concrete network of our proposed model can be derived 
        based on the process that minimizes its energy function. 

  \item Our network has a ladder-shaped architecture 
        with internal and external spaces, 
        each of which has a pair of a feedback path and an input port. 

  \item Some groups of synaptic connections exist in the whole network. 
        Among them, only those in layered networks with static neurons have plasticity. 
        All other connections are fixed. 

  \item Two kinds of dynamics, activity dynamics for dynamical neurons 
        and learning dynamics for synaptic connections in a layered network, 
        progress simultaneously, 
        and both cooperatively determine the network's total behavior. 

  \item We considered our dynamical neural network a model 
        and supposed two modes, Learning and Association (Types \#0 or \#1), 
        which depend on whether a pair of a feedback path and an input port 
        is in the internal or external space or in both. 
        In either mode, the total dynamics are produced 
        due to the cooperation between dynamic and static neurons. 

\end{itemize}

\vspace{2mm} 
\noindent 
\( \diamondsuit \) Learning Mode

\begin{itemize}

  \setlength{\itemsep}{-0.05mm}

  \item When appropriate sinusoidal and/or quasi-sinusoidal waves 
        are applied in the Learning Mode 
        from the outside of the model to the input ports, 
        the network can automatically acquire the corresponding mapping relationship 
        in {\sl Internetwork} connecting the internal space with the external space.
        The sinusoidal signal here corresponds to repetitive neuronal bursting, 
        which represents a periodic wave 
        in the short-term average density of nerve impulses 
        or in the membrane potential. 

  \item When the given periodic signals have different waveforms, 
        a non-linear mapping relationship (instead of a linear one) 
        is formed in {\sl Internetwork}, 
        depending on the shapes of the waves. 

  \item The model can be expanded to a two-dimensional scheme. 
        When sinusoidal waves with adequately different frequencies 
        are applied to the input ports, 
        a two-dimensional mapping relationship can be acquired 
        in {\sl Internetwork} based on its generalization capability. 

  \item Such a situation is recommended where the frequencies 
        of these two sinusoidal waves are 
        either only slightly different or extremely different; 
        this aspect is coincident with the framework of Lissajous curves.

  \item When periodic signals with different waveforms are applied 
        to the internal and external spaces, 
        non-linear mapping relationships can be acquired in {\sl Internetwork} 
        even in a two-dimensional model. 

  \item In the Learning Mode, we can confirm some interesting phenomena 
        in which two kinds of dynamics, that is, 
        activity dynamics due to dynamical neurons
        and learning dynamics in a layered network with static neurons, 
        seem to be cooperatively related. 

\end{itemize}

\vspace{2mm} 
\noindent 
\( \diamondsuit \) Association Mode

\begin{itemize}

  \setlength{\itemsep}{-0.05mm}

  \item In the Association Mode, we found that 
        the speed of convergence from an initial point to a goal point varies 
        depending on how the mapping relationship 
        between the internal and external spaces is warped. 
        In the two-dimensional model, 
        we confirmed that a convergence trajectory can become out of the straight 
        due to the warping of the mapping relationship. 

  \item In Type \#1 of the Association Mode 
        with a pair of a feedback path and an input port 
        only in the external space, 
        we clarified that a convergence trajectory curves more greatly  
        compared with Type \#0 of the Association Mode 
        with a pair of a feedback path and an input port 
        only in the internal space. 

  \item If {\sl Internetwork} were trained to be linear through its full range, 
        the dynamics for the internal and external spaces in both Type \#0 and Type \#1 
        of the Association Mode would be identical 
        to those of the original Hopfield model. 

  \item Although we introduced the Association Mode in Section 4 
        and the Constrained Association Mode in Section 5, 
        it is also reasonable 
        to call the former the ``Unconstrained'' Association Mode. 

\end{itemize}

\vspace{2mm} 
\noindent 
\( \diamondsuit \) Inverse Mapping

\begin{itemize} 

  \setlength{\itemsep}{-0.05mm}

  \item In Type \#0 of the Constrained Association Mode, 
        no signal propagates through {\sl Backward Subnet} in principle.
        Applying a very slow target trajectory (instead of a goal point) 
        to the input port in Type \#1 of the Constrained Association Mode, 
        signals flowing through {\sl Backward Subnet} become quite small 
        due to the function of the dynamical neurons in the internal space. 
        Hence, in both architectures of the Constrained Association Mode 
        with an extremely slow target trajectory, 
        curving for the output trajectories in the internal and external spaces 
        is generated only by the static mapping relationship of {\sl Internetwork}. 

  \item In Type \#1 of the Constrained Association Mode 
        with a pair of a feedback path and an input port 
        only in the external space, 
        if \( M \) is the mapping relationship of {\sl Forward Subnet}, 
        a phenomenon appears in such a manner that 
        a mapper with \( M^{-1} \) does exist, 
        although it does not exist explicitly in the whole network. 
        In other words, an output trajectory is generated in the internal space, 
        that is transformed by a mapper with \( M^{-1} \) 
        from a target trajectory applied in the external space. 

\end{itemize}

\vspace{2mm} 
Although the Learning Mode and the two kinds of 
Association Modes (Unconstrained / Constrained) were 
artificially classified on the basis of their structural differences 
and the ON/OFF operation of the learning dynamics given by Eq. (29) 
was consistently selected here, 
it might be possible to construct a model 
in which such modes can be switched automatically 
according to the situation and the form of the input signals 
given from the outside. 
In the Learning Mode discussed in Section 3, 
a non-distorted sinusoidal wave was applied 
to the input port in the internal space, 
but the waveform is not limited to such a case. 
For the two-dimensional model, 
the two inputs for the horizontal and vertical axes have no correlation, 
but it is quite significant to examine various cases 
in which the two inputs are correlated. 
This paper refers only to one-dimensional and two-dimensional models. 
After increasing the dimension from one to two, 
it should be straightforward to expand it to three and higher. 
Also, we only treated convergent dynamics toward a fixed point 
throughout our paper, but we can discuss divergent dynamics 
by changing the signs of the terms in the energy function
\cite{Tsutsumi1999a}\cite{Tsutsumi1999b}\cite{Tsutsumi2001}.
These expansions will be developed in future work. 

Even though, as suggested previously, 
our primary motivation is not the solution of any practival problems, 
we expect that our framework, with a dynamical system relaxing in a warped space, 
has the potential to several practical applications. 
Examples are advanced associative memory tasks, optimization problems, 
or control issues in which each constraint must be minimized 
in the corresponding non-linear subspace. 
If the mapping relationship of {\sl Internetwork} is chosen appropriately, 
it may be possible to expand the basins of attraction in a warped space. 
Our model then might improve the performance of associative recall 
and enhance the memory capacity when applying it 
to an associative memory task
\cite{Cabrera2016}.
In those cases, as discussed in detail in Section 5, 
it is crucial to provide an adequate constraint 
in the form of input signals from the outside to the model 
in order to control the convergent or divergent trajectory 
in the internal space as intended. 
In this paper, we mainly discussed depth-wise or vertical (hierarchical) 
modularity, but not spread-wise or horizontal modularity as in internetworking. 
This point is significant not only from an architectural viewpoint 
but also in the sense of illustrating the performance of our proposed model. 
Our approach opens a large array of subjects. 
In future work, we plan to discuss a general model 
that is not limited to internal and external spaces 
but contains multiple subspaces.

% References
\bibliographystyle{acm}
\bibliography{tsutsumi_niebur_01_references}

% Acknowledgements
\vspace{2mm}
\noindent
{\bf Acknowledgements:} This work was supported in part 
by Ryukoku University Fund for Overseas Research, 
Office of Naval Research Grant N00014-22-1-2699, 
and National Science Foundation grant 1835202.

\end{document}